\documentclass[preprint,aps,eqsecnum,showpacs]{revtex4}
\usepackage{graphicx}
\usepackage{amssymb,amsmath}

\begin{document}

\title{A quantum mechanical model of the Riemann zeros 
}

\author{Germ\'an Sierra}

\affiliation{Instituto de F\'{\i}sica Te\'orica, CSIC-UAM, Madrid, Spain}
%\date{December, 2007}

% started the 5 June 2007

\bigskip\bigskip\bigskip\bigskip

%%%%%%%%%%%%%%% MATH CHARACTERS %%%%%%%%%%%%%%%%%%%%%%%%%%%%
%
\font\numbers=cmss12
%\font\numbers=cmu10 scaled\magstep1
\font\upright=cmu10 scaled\magstep1
\def\stroke{\vrule height8pt width0.4pt depth-0.1pt}
\def\topfleck{\vrule height8pt width0.5pt depth-5.9pt}
\def\botfleck{\vrule height2pt width0.5pt depth0.1pt}
\def\Zmath{\vcenter{\hbox{\numbers\rlap{\rlap{Z}\kern
0.8pt\topfleck}\kern 2.2pt
                   \rlap Z\kern 6pt\botfleck\kern 1pt}}}
\def\Qmath{\vcenter{\hbox{\upright\rlap{\rlap{Q}\kern
                   3.8pt\stroke}\phantom{Q}}}}
\def\Nmath{\vcenter{\hbox{\upright\rlap{I}\kern 1.7pt N}}}
\def\Cmath{\vcenter{\hbox{\upright\rlap{\rlap{C}\kern
                   3.8pt\stroke}\phantom{C}}}}
\def\Rmath{\vcenter{\hbox{\upright\rlap{I}\kern 1.7pt R}}}
\def\Z{\ifmmode\Zmath\else$\Zmath$\fi}
\def\Q{\ifmmode\Qmath\else$\Qmath$\fi}
\def\N{\ifmmode\Nmath\else$\Nmath$\fi}
\def\C{\ifmmode\Cmath\else$\Cmath$\fi}
\def\R{\ifmmode\Rmath\else$\Rmath$\fi}
\def\H{{\cal H}}
\def\NN{{\cal N}}
\def\cl{{\rm cl}}
\def\RS{{\rm RS}}
\def\E{{\rm E}}

\begin{abstract}
In 1999 Berry and Keating showed that a regularization of the 
1D classical Hamiltonian $H = xp$ gives semiclassically
the smooth counting function of the Riemann zeros. 
In this paper we first generalize this 
result by considering a phase space delimited by two
boundary functions in position and momenta, which 
induce a fluctuation term in the counting of energy
levels. We next quantize the $x p$ Hamiltonian, adding an
interaction term that depends on two wave functions
associated to the classical boundaries in phase space. 
The general model is solved exactly, obtaining a continuum 
spectrum with discrete bound states embbeded in it. 
We find the boundary wave functions, associated
to the Berry-Keating regularization, for which the
average Riemann zeros become resonances. A spectral
realization of the Riemann zeros is achieved 
exploiting the symmetry of the model under the 
exchange of position and momenta which is 
related to the duality symmetry of the zeta
function. The boundary wave functions, giving
rise to the Riemann zeros, are found using the
Riemann-Siegel formula of the zeta function. 
Other Dirichlet L-functions are shown to find a natural
realization in the model.  
\end{abstract}


\maketitle

\vskip 0.2cm

%       DEFINITIONS FOR TEX
%
%%%%%%%%%%%%%%%%%%%%%%%%%%%%%%%%%%%%%%%%%%%%%%%%%%%%%%%%%%%%%%%
%
%
%\def\e{\'e}
%\def\ee{\`e}
%%%%%%%%%%%%%%%%%%%DEFINITIONS%%%%%%%%%%%%%%%%%%%%%%%%%%%%%%%%%
%
\def\oti{{\otimes}}
\def\lb{ \left[ }
\def\rb{ \right]  }
\def\tilde{\widetilde}
\def\bar{\overline}
\def\hat{\widehat}
\def\*{\star}
\def\[{\left[}
\def\]{\right]}
\def\({\left(}      \def\BL{\Bigr(}
\def\){\right)}     \def\BR{\Bigr)}

%%%%%%%%%%%%%%%%%%%%%%%%%%%%%%%%%%%%%%%%%%%%%%%%%%%%%%%%%%%%%%%
%
\def\zb{{\bar{z} }}
\def\zbar{{\bar{z} }}
\def\frac#1#2{{#1 \over #2}}
\def\inv#1{{1 \over #1}}
\def\half{{1 \over 2}}
\def\d{\partial}
\def\der#1{{\partial \over \partial #1}}
\def\dd#1#2{{\partial #1 \over \partial #2}}
\def\vev#1{\langle #1 \rangle}
\def\ket#1{ | #1 \rangle}
\def\rvac{\hbox{$\vert 0\rangle$}}
\def\lvac{\hbox{$\langle 0 \vert $}}
\def\2pi{\hbox{$2\pi i$}}
\def\e#1{{\rm e}^{^{\textstyle #1}}}
\def\grad#1{\,\nabla\!_{{#1}}\,}
\def\dsl{\raise.15ex\hbox{/}\kern-.57em\partial}
\def\Dsl{\,\raise.15ex\hbox{/}\mkern-.13.5mu D}
%
%%%%%%%%%%%%%%%%%%%%GREEK LETTERS%%%%%%%%%%%%%%%%%%%%%%%%%%%%%%
%
%\def\th{\theta}        \def\Th{\Theta}
\def\ga{\gamma}     \def\Ga{\Gamma}
\def\be{\beta}
\def\al{\alpha}
\def\ep{\epsilon}
\def\vep{\varepsilon}
\def\dep{d}
\def\arc{{\rm Arctan}}
\def\la{\lambda}    \def\La{\Lambda}
\def\de{\delta}     \def\De{\Delta}
\def\om{\omega}     \def\Om{\Omega}
\def\sig{\sigma}    \def\Sig{\Sigma}
\def\vphi{\varphi}
\def\sign{{\rm sign}}
\def\he{\hat{e}}
\def\hf{\hat{f}}
\def\hg{\hat{g}}
\def\ha{\hat{a}}
\def\hb{\hat{b}}
\def\hx{\hat{x}}
\def\hp{\hat{p}}
\def\f{{\bf f}}
\def\g{{\bf g}}
\def\a{{\bf a}}
\def\b{{\bf b}}
\def\fl{{\rm fl}}
\def\sm{{\rm sm}}
\def\QM{{\rm QM}}

%
%%%%%%%%%%%%%%%%%%%CALIGRAPHIC LETTERS%%%%%%%%%%%%%%%%%%%%%%%%%
%
\def\CA{{\cal A}}   \def\CB{{\cal B}}   \def\CC{{\cal C}}
\def\CD{{\cal D}}   \def\CE{{\cal E}}   \def\CF{{\cal F}}
\def\CG{{\cal G}}   \def\CH{{\cal H}}   \def\CI{{\cal J}}
\def\CJ{{\cal J}}   \def\CK{{\cal K}}   \def\CL{{\cal L}}
\def\CM{{\cal M}}   \def\CN{{\cal N}}   \def\CO{{\cal O}}
\def\CP{{\cal P}}   \def\CQ{{\cal Q}}   \def\CR{{\cal R}}
\def\CS{{\cal S}}   \def\CT{{\cal T}}   \def\CU{{\cal U}}
\def\CV{{\cal V}}   \def\CW{{\cal W}}   \def\CX{{\cal X}}
\def\CY{{\cal Y}}   \def\CZ{{\cal Z}}

\def\Hp{{\mathbb{H}^2_+}} 
\def\Hm{{\mathbb{H}^2_-}}

\def\rvac{\hbox{$\vert 0\rangle$}}
\def\lvac{\hbox{$\langle 0 \vert $}}
\def\comm#1#2{ \BBL\ #1\ ,\ #2 \BBR }
\def\2pi{\hbox{$2\pi i$}}
\def\e#1{{\rm e}^{^{\textstyle #1}}}
\def\grad#1{\,\nabla\!_{{#1}}\,}
\def\dsl{\raise.15ex\hbox{/}\kern-.57em\partial}
\def\Dsl{\,\raise.15ex\hbox{/}\mkern-.13.5mu D}
%
%%%%%%%%%%%%%%%%%%%%GREEK LETTERS%%%%%%%%%%%%%%%%%%%%%%%%%%%%%%
%
%%%%%%%%%%%%%%% MATH CHARACTERS %%%%%%%%%%%%%%%%%%%%%%%%%%%%
%
\font\numbers=cmss12
%\font\numbers=cmu10 scaled\magstep1
\font\upright=cmu10 scaled\magstep1
\def\stroke{\vrule height8pt width0.4pt depth-0.1pt}
\def\topfleck{\vrule height8pt width0.5pt depth-5.9pt}
\def\botfleck{\vrule height2pt width0.5pt depth0.1pt}
\def\Zmath{\vcenter{\hbox{\numbers\rlap{\rlap{Z}\kern
0.8pt\topfleck}\kern 2.2pt
                   \rlap Z\kern 6pt\botfleck\kern 1pt}}}
\def\Qmath{\vcenter{\hbox{\upright\rlap{\rlap{Q}\kern
                   3.8pt\stroke}\phantom{Q}}}}
\def\Nmath{\vcenter{\hbox{\upright\rlap{I}\kern 1.7pt N}}}
\def\Cmath{\vcenter{\hbox{\upright\rlap{\rlap{C}\kern
                   3.8pt\stroke}\phantom{C}}}}
\def\Rmath{\vcenter{\hbox{\upright\rlap{I}\kern 1.7pt R}}}
\def\Z{\ifmmode\Zmath\else$\Zmath$\fi}
\def\Q{\ifmmode\Qmath\else$\Qmath$\fi}
\def\N{\ifmmode\Nmath\else$\Nmath$\fi}
\def\C{\ifmmode\Cmath\else$\Cmath$\fi}
\def\R{\ifmmode\Rmath\else$\Rmath$\fi}
%%%%%%%%%%%%%%%%%%%%%%%%%%%%%%%%%%%%%%%%%%%%%%%%%%%%%%%%%%%%%%%%%
 %%%%%%%%%%%%%%%%%% END OF DEFINITIONS %%%%%%%%%%%%%%%%%%%%%%
 %%%%%%%%%%%%%%%%%%%%%%%%%%%%%%%%%%%%%%%%%%%%%%%%%

\def\barray{\begin{eqnarray}}
\def\earray{\end{eqnarray}}
\def\beq{\begin{equation}}
\def\eeq{\end{equation}}

\def\no{\noindent}

\def\s{\sigma}
\def\Ga{\Gamma}

\def\L{{\cal L}}
\def\g{{\bf g}}
\def\K{{\cal K}}
\def\I{{\cal I}}
\def\M{{\cal M}}
\def\F{{\cal F}}

\def\Im{{\rm Im}}
\def\Re{{\rm Re}}
\def\ti{{\tilde{\phi}}}
\def\tR{{\tilde{R}}}
\def\tS{{\tilde{S}}}
\def\tF{{\tilde{\cal F}}}

\section{Introduction}

At the beginning of the XX century Polya and Hilbert made
the bold conjecture that the imaginary part of the Riemann
zeros could be the oscillation frequencies of a physical system.
If true this suggestion would 
imply a proof of the celebrated Riemann hypothesis (RH). 
The importance of this conjecture lies in its connection
with the prime numbers. If the RH is true then 
the statistical distribution of the primes will
be constrained in the most favorable way \cite{Edwards,Titchmarsh2}.
Otherwise, in  
the words of Bombieri, the
failure of the RH would create havoc in the 
distribution of the prime numbers \cite{Bombieri}
(see also \cite{Sarnak,Conrey,Watkins,Rosu,Elizalde} for reviews on the RH). 

After the advent of Quantum Mechanics, 
the Polya-Hilbert conjecture was formulated
as the existence of a self-adjoint operator whose
spectrum contains the imaginary part of the 
Riemann zeros. This conjecture was for a long time
regarded as a wild speculation until 
the works of Selberg in the 50's and those of 
Montgomery in the 70's. Selberg 
found a remarkable duality between 
the length of geodesics on a Riemann
surface and the  eigenvalues of the 
 Laplacian operator defined on it \cite{Selberg}. This 
duality is encapsulated in the so called
Selberg trace formula, which has a strong similarity
with the Riemann explicit formula
relating the zeros and the prime numbers. 
The Riemann zeros would correspond
to the eigenvalues, and the primes to the 
geodesics. This classical versus quantum 
version of the primes and the zeros is also at the heart of 
the so called  Quantum Chaos approach to the RH. 

Quite independently
of Selberg«s  work, Montgomery showed that the
Riemann zeros are distributed randomly and 
obeying locally the statistical law of the 
Random Matrix Theory  (RMT) \cite{Mont}. 
The RMT was originally proposed
to explain the chaotic behaviour of the spectra of nuclei
but it has applications in another branches
of Physics, specially in  Condensed Matter \cite{Mehta}. 
There are several universality  classes
of random matrices, and it turns out that
the one related to the 
Riemann zeros is the gaussian unitary ensemble (GUE)
associated to random hermitean matrices. 
Montgomery analytical results  found an impressive
numerical confirmation in the works of
Odlyzko in the 80's, so that the GUE 
law, as applied  to the Riemann zeros 
is nowadays called the Montgomery-Odlyzko law \cite{Odl}. 
An important hint  suggested by this  law
is that the Polya-Hilbert Hamiltonian $H$ 
must break the time reversal symmetry. The reason being that
the GUE statistics describes random Hamiltonians
where this symmetry is broken. A simple example 
is provided by materials with impurities 
 subject to an external magnetic field,
as in the Quantum Hall effect.  

A further step in the Polya-Hilbert-Montgomery-Odlyzko
pathway was taken by Berry \cite{B-chaos,Berry1}. 
who noticed a similarity between the formula
yielding the  fluctuations of the number of zeros, 
around its average position $E_n \sim 2 \pi n/ \log n$,
and a formula giving the fluctuations of the
energy levels of a Hamiltonian obtained
by the quantization of a classical chaotic system \cite{Gutzwiller}. 
The comparison between these two formulas suggests
that the prime numbers $p$  correspond to the isolated
periodic orbits whose period is $\log p$. In the
Quantum Chaos scenario the prime numbers appear
as classical objects, while the Riemann zeros are
quantal. This classical/quantum interpretation
of the primes/zeros is certainly reminiscent
of the one underlying the Selberg trace formula
mentioned earlier. A success of the 
Quantum Chaos approach is that it explains the
deviations from the GUE law of the zeros found
numerically by Odlyzko. The similarity between
the fluctuation formulas 
described above, while rather appealing, has
a serious drawback observed by Connes which
has to do with an overall sign difference between them \cite{Connes}. 
It is as if the periodic orbits were missing 
in the underlying classical chaotic dynamics, a fact that
is difficult to understand physically. This and other
observations lead  Connes to propose an abstract
approach to the RH based on discrete mathematical
objects known as adeles \cite{Connes}. 
The final outcome of Connes
work is a trace formula whose proof, not yet found, 
amounts to that of a generalized version of the RH. 
In Connes approach there is an operator, which plays
the role of the Hamiltonian, whose spectrum
is a continuum with missing spectral lines
corresponding to the Riemann zeros. 
We are thus confronted with two possible physical
realizations of the Riemann zeros, either
as point like spectra or as missing spectra
in a continuum. Later on we shall see that 
both pictures can be reconciled in a QM 
model having a discrete spectra embedded in a 
continuum. 

The next step within the Polya-Hilbert framework
came in 1999 when  Berry and Keating \cite{BK1,BK2} on one
hand and Connes  \cite{Connes} on the other, proposed that the
classical Hamiltonian $H = x p$, where $x$
and $p$ are the position and momenta of a 1D particle,
is closely related to the Riemann zeros. 
This striking suggestion
was based on a semiclassical analysis 
of $H = x p$, which led these authors to reach 
quite opposite
conclusions regarding the possible spectral
interpretation of the Riemann zeros. The origin
of the disagreement is due to the choice
of different regularizations of $H = xp$.
Berry and Keating choosed a Planck cell
regularization in which case the smooth
part of the Riemann zeros appears semiclassically 
as discrete energy levels. Connes, on the other
hand choosed an upper cutoff for the position
and momenta
which gives semiclassically a continuum 
spectrum where the smooth zeros are missing. 
All these semiclassical results  are heuristic
and lack so far of a consistent quantum version. 
It is the aim of this paper to provide such a
quantum version in the hope that it will sed
new light concerning the spectral realization
of the Riemann zeros.

The organization of the paper is as follows. 
In section II we review the semiclassical approaches
to $H = xp$ due to Berry, Keating and Connes which give
an heuristic derivation of the asymptotic behaviour of the smooth
part of the Riemann zeros. Then, we generalize the semiclassical Berry-Keating
Planck cell regularization of $xp$  by means  of two classical functions which 
define a {\em wiggly } boundary for the allowed semiclassical region
in phase space. This generalization allow us to 
explain semiclassically the fluctuation term in the spectrum. 
In section III  we define  the quantum Hamiltonian associated to the 
 semiclassical approach introduced  above. The Hamiltonian 
 is given by the quantization of $H= xp$ plus an interaction
 term that depends on two generic boundary wave functions associated
 to the classical boundary functions of the semiclassical approach. 
 In section IV  we solve the Schroedinger equation finding the 
 exact eigenfunctions and eigenenergies in terms of a function
 $\F(E)$ which plays the role of a Jost function for this model, and 
 whose analyticity properties are studied in section V. In section VI
 we find  the boundary wave functions that give rise to the quantum
 version of the semiclassical Berry-Keating model for the 
 smooth zeros of the Riemann zeta function, which 
 are common to all the even Dirichlet L-functions. 
We also find the boundary wave functions associated to the smooth
 approximation of the zeros of the odd Dirichlet L-functions. 
 In section VII we quantize the relation 
 between the fluctuation part of the spectrum 
and the semiclassical phase boundaries, 
obtaining the equations satisfied  by the boundary wave functions, and we
solve them  explicitely. Finally, using the duality properties of 
these wave functions 
and the Riemann-Siegel formula of the zeta function we 
find a model whose Jost function is proportional to the zeta
function. From this fact, and making some additional asumptions, 
we show that the Riemann zeros on the critical
line are bound states of the model. However we cannot exclude
the existence of zeros outside the critical line, which 
would imply a proof of the RH. 
We describe in an appendix the computation
of the wave functions associated to the smooth and exact Riemann zeros.

The present work is closely related to those in references 
\cite{Sierra1,Sierra2,Sierra3}, 
where we studied an interacting version of the $xp$ Hamiltonian based on the 
relation of this model with the so called Russian doll model
of superconductivity \cite{RD1,RD2,links}. For a field theoretical 
approach to the RH inspired
by the latter works see reference \cite{Andre-RH}. We would like also
 to mention
some important differences between the present paper and those of references
\cite{Sierra1,Sierra2,Sierra3}. 
First of all,  the position variable $x$ was choosen in 
\cite{Sierra1,Sierra2,Sierra3}
 to belong to the finite interval $(1,N)$ with $N \rightarrow \infty$, while
in this paper we choose the half line $(0, \infty)$ which gives a more
symmetric treatment between the position and momentum variables. 
Secondly,   in the earlier references the interaction term was added
to the inverse Hamiltonian $1/(xp)$, while in this paper we add
the interaction directly  to the Hamiltonian $xp$, which is more
natural from a physical viewpoint. We have also tried 
to make an extensive use of the duality symmetry of the Riemann
zeta function reflected in the functional relation it satisfies.

\section{Semiclassical approach}

The classical Berry-Keating-Connes (BKC) Hamiltonian  
\cite{BK1,BK2,Connes} 
\beq
H^{\rm cl}_{0} = x \;  p,  
\label{s1}
\eeq
\no 
has classical trayectories  given by the  hyperbolas 
(see fig.1a)
\beq
x(t) = x_0 \; e^{t} , \quad p(t) = p_0 \;  e^{-t}.  
\label{s2}
\eeq
\begin{figure}[t!]
\begin{center}
\includegraphics[height= 7.0 cm,angle= 0]{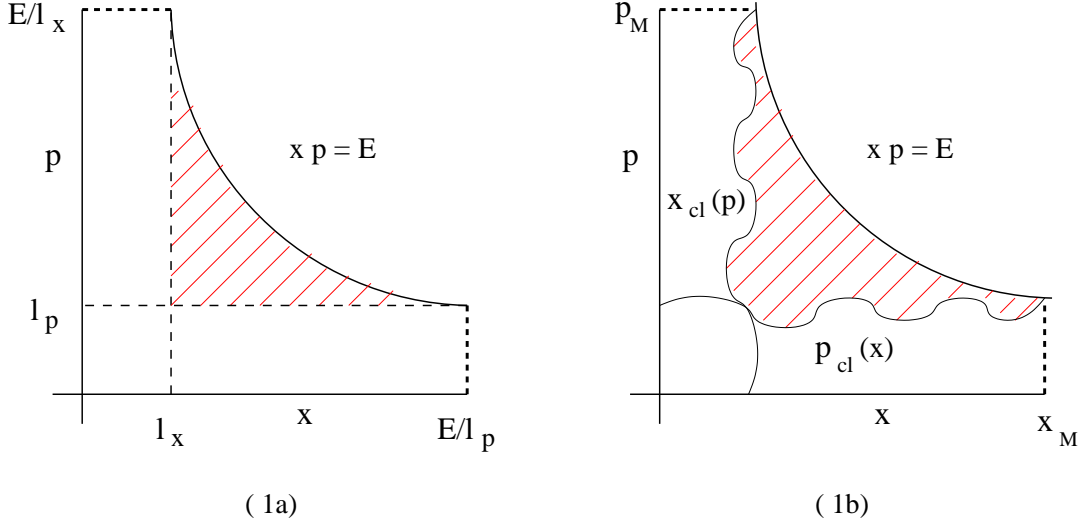}
\end{center}
\caption{
1a) a classical trayectory (\ref{s2}).  
The region in shadow is the allowed 
phase space of the semiclassical 
regularizations of Berry and Keating.
1b) generalization of the phase space
region given by equations (\ref{s10})
}
\label{boundaries}
\end{figure}
\no
The dynamics is unbounded, so 
one should not expect a discrete spectrum 
even at the semiclassical level. 
To overcome this difficulty, Berry and Keating proposed
in 1999 to restrict the phase space of the $xp$ model
to those points $(x,p)$ 
where  $|x| > l_x$ and $|p| > l_p$, 
with $l_x \, l_p = 2 \pi \hbar$. 
These constraints lead to a finite 
number of semiclassical states, $\CN(E)$, with energy 
between 0 and $E$ given by
\beq
\CN(E) = \frac{A}{2 \pi \hbar}, 
\label{s3}
\eeq
\no
where $A$ is the area of the allowed phase space region 
below the curve $E = x p$. The result, in units $\hbar = 1$, is 
\beq
\CN_{BK}(E) =  \frac{E}{2 \pi} 
\left( \log \frac{E}{2 \pi} -1 \right) + 1
\label{s4} 
\eeq
\no 
which agrees with the 
asymptotic limit of the smooth part of the formula
giving the 
number of  Riemann zeros whose imaginary part lies
in the interval $(0,E)$, 
\beq
\langle \CN(E) \rangle \sim  \frac{E}{2 \pi} 
\left( \log \frac{E}{2 \pi} -1 \right) + \frac{7}{8} + O(E^{-1}).
\label{s5} 
\eeq
\no
The exact formula for the number of zeros, $\CN_R(E)$, due to Riemann,
also contains a fluctuation term which depends on 
the zeta function \cite{Edwards} (see fig.\ref{n-formula}), 
\barray
\CN_R(E) & = & \langle \CN(E) \rangle  + \CN_{\rm fl}(E)  
\label{s6}
\\
\langle \CN(E) \rangle &  = & 
\frac{\theta(E)}{\pi}  + 1 
\nonumber 
 \\
\CN_{\rm fl}(E) & = &   \frac{1}{\pi} \Im \log \zeta \left( 
\frac{1}{2} + i E \right) \nonumber 
\earray
\no
where $\theta(E)$ is the phase of the Riemann zeta 
function $\zeta(1/2 - i E)$, 
\beq
\theta(E) = \Im \log \Gamma \left( \frac{1}{4} + \frac{i}{2} E \right) 
- \frac{E}{2} \log \pi 
\label{s7}
\eeq
\no
whose asymptotic expansion 
\beq
\theta(E) = \frac{E}{2} \log \left( \frac{E}{2 \pi} \right)  
- \frac{E}{2} - \frac{\pi}{8} + O(E^{-1}) 
\label{s7-1}
\eeq
\no 
 yields (\ref{s5}). 
The function $\zeta(s)$, for $\Re \;  s > 1$, can be related
to the prime numbers $p$ thanks to the Euler
product formula
\beq
\zeta(s) = \prod_{p > 1} \frac{1}{1 - p^{-s}},\qquad
\Re \;  s > 1 
\label{s7-2}
\eeq
\no
\begin{figure}[t!]
\begin{center}
\includegraphics[height= 7.0 cm,angle= 0]{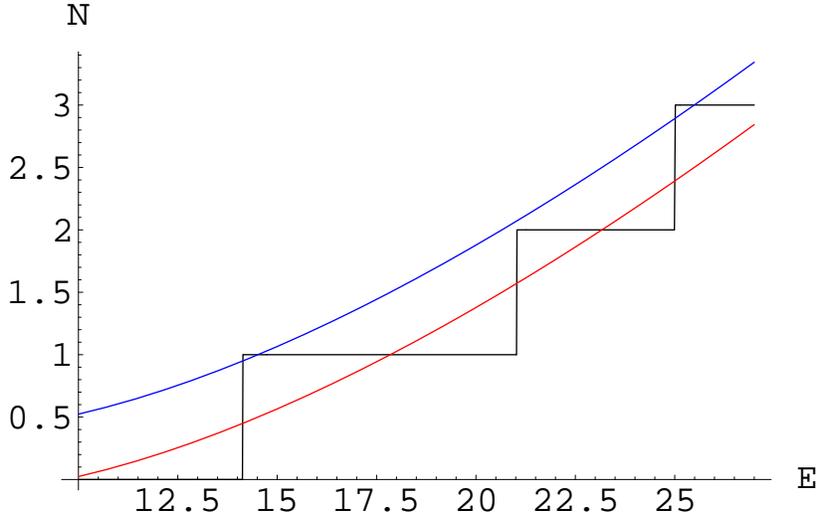}
\end{center}
\caption{
Number of Riemann zeros in the interval $(0,E)$:
black: exact formula (\ref{s6}), red: smooth function
$\langle \CN(E) \rangle$,   blue: $\langle \CN(E) \rangle + 1/2$.
}
\label{n-formula}
\end{figure}
\no
This expression diverges if $\Re \; s = 1/2$, however
one can heuristically use it to write the
fluctuation term in  (\ref{s6}) as
\beq
\CN_{\rm fl}(E) = - \frac{1}{\pi} 
\sum_{p} \sum_{m=1}^\infty  
\frac{1}{m \; p^{m/2}} \; \sin ( m E \log p) 
\label{s8}
\eeq
\no
which gives a reasonable result after 
truncating the sum over the primes. 
As observed by Berry, eq.(\ref{s8}) 
resembles formally the fluctuation part
of the spectrum 
of a classical 1D chaotic Hamiltonian 
with isolated periodic orbits 
\beq
\CN_{\rm fl}(E) =  \frac{1}{\pi} 
\sum_{\gamma_p} \sum_{m=1}^\infty  
\frac{1}{m \; 2 \; {\rm sinh} (m \lambda_p/2)} \; \sin (S_{\rm cl}(E)) 
\label{s9}
\eeq
\no
where $\gamma_p$ denotes the primitive periodic
orbits, the label $m$ describes the windings of those orbits,
$ \pm \lambda_p$ are the instability exponents and $S_{\rm cl}(E)$
is the classical action, which is equal to $m E T_{\gamma_p}$, 
with $ T_{\gamma_p}$
the period of $\gamma_p$. Comparing (\ref{s8})
and (\ref{s9}), Berry conjectured the existence
of a classical chaotic Hamiltonian whose
primitive periodic orbits would be labelled 
by the prime numbers $p=2,3, \dots$, with periods
$T_p= \log p$ and instability exponents
$\lambda_p = \pm \log p$ \cite{B-chaos,Berry1}. Moreover, since each
orbit is counted once, the Hamiltonian
must break time reversal (otherwise there would be a factor
$2/\pi$ in front of eq. (\ref{s8}) instead of $1/\pi$).
The quantization of this classical chaotic Hamiltonian
would likely contain the Riemann zeros in its spectrum.
This idea is the key of the Quantum Chaos
approach to the Riemann hypothesis. 

Besides the
fact that the earlier Hamiltonian
has not yet been found there is the Connes
criticism that the similarity between eqs.(\ref{s8})
and (\ref{s9}) fails in two issues. The first is the
 overall minus sign in (\ref{s8})
as compared to (\ref{s9}), and  the second 
is that the term $2 \; {\rm sinh}(m \lambda_p/2)$ 
only becomes $p^{m/2}$ when $m \rightarrow \infty$. 
Connes relates the {\em minus sign} problem
to an alternative interpretation
of the Riemann zeros as missing spectral lines
as opposed to the conventional one (we shall come back
later to these conflicting  interpretations).
These two problems were
the main  Connes's motivations to develop the adelic
approach to the RH.

As we saw above, the Quantum Chaos approach
suggests that the fluctuation part of the
spectrum 
of the yet unknown Riemann Hamiltonian 
has a classical origin 
related to the prime numbers. Taking into account
the  Berry-Keating
heuristic derivation of the smooth part of the spectrum,
it is tempting to extend the semiclassical
approach in order to explain the fluctuation term in the Riemann formula
for the zeros.
The simplest idea is to generalize the allowed
phase space of the $xp$  Hamiltonian replacing 
the boundaries $|x| = l_x$ and $|p| = l_p$
by two curves $x_\cl(p)$ and 
$p_\cl(x)$, such that (see fig 1b)
\beq
x > x_{\cl}(p), \qquad |p| > p_\cl(x)  
\label{s10}
\eeq
\no
where   $x_\cl(p)$ and 
$p_\cl(x)$, are positive functions satisfying
\barray 
x_{\cl}(p) =  x_\cl(-p)& > 0, & \qquad \forall \; p \in \Rmath 
\label{s11} \\
p_\cl(x) & >0 &   \qquad \forall \;  x \in \Rmath_+  
\nonumber 
\earray 
\no
These conditions split the allowed phase space
into two disconnected regions in the first and
forth quadrants of the $x p$ plane. Notice that 
$x$ is always positive while $p$ can be either
positive or negative. 
The BK boundaries obviously correspond to the choice
\beq
{\rm BK} : \;  x_{\cl}(p)= l_x, \qquad p_\cl(x) = l_p   
\label{s12}
\eeq
\no
For the extended BC's  the minimal distance
$l_x$ and minimal  momentum $l_p$ 
can be defined as the intersection point
of the curves, $x_\cl(p)$ and 
$p_\cl(x)$, which we shall assume to be unique,
and satisfying 
\beq
x_{\cl}(l_p)= l_x, \qquad p_\cl(l_x) = l_p   
\label{s13}
\eeq
\no
The classical $x p$ Hamiltonian together with the BK conditions
have the exchange symmetry
\beq
\frac{x}{l_x} \leftrightarrow \frac{p}{l_p} 
\label{s14}
\eeq
\no
whose generalization to the extended model is 
\beq
\frac{x_\cl( l_p x/l_x)}{l_x} =  \frac{p_\cl(x)}{l_p} 
\label{s15}
\eeq
\no
The counting of semiclassical states is based
again on eq. (\ref{s3}). The 
area below the curve $E =  x p$ and bounded
by the conditions (\ref{s10}) is given by 
(see fig.1b) 
\barray 
A & = &  \int_{l_x}^{x_I} dx  \int_{p_\cl(x)}^{l_p x/l_x}
dp + \int_{x_I}^{x_M} d x  \int_{p_\cl(x)}^{E/x}  dp  
\label{s16} \\
&+ & \int_{l_p}^{p_I} dp \int_{x_\cl(p)}^{l_x p/l_p} dx
+ \int_{p_I}^{p_M} dp \int_{x_{\cl}(p)}^{E/p} dx 
\nonumber 
\earray
\no
The quantities $x_M$, $p_M$ (resp. $x_I, p_I$) 
are the position and momenta of the points
where the curve $E = x p$ intersects the boundaries
 $p_\cl(x)$,  $x_\cl(p)$ (resp. the line $x/l_x = p/l_p$),
and satisfy,
\beq
E = x_M \;  p_\cl(x_M) = x_\cl(p_M) \;  p_M = x_I p_I, \quad
\frac{x_I}{l_x} = \frac{p_I}{l_p} 
\label{s17}
\eeq
\no
The integration of (\ref{s16}) yields
\barray 
A & = & E \log \left(  \frac{E}{ l_x  l_p} \right)  + E - l_x l_p 
\label{s18} \\
& - & E \log \left( \frac{ p_\cl(x_M)}{ l_p} \right) 
-  E \log \left( \frac{ x_\cl(p_M)}{l_x}  \right) 
\nonumber \\
& - &  \int_{l_x}^{x_M} dx \;  p_\cl(x) -
\int_{l_p}^{p_M} dp \;  x_\cl(p) 
\nonumber    
\earray
\no
Partial integrating the last two terms
in (\ref{s18}) and dividing by $h = l_x l_p = 2 \pi (\hbar =1)$,
the semiclassical value of $\CN(E)$ reads
\barray 
\CN(E) & = & \CN_{BK}(E) 
\label{s19} \\
& - & \frac{E}{2 \pi}  \log \left( \frac{ p_\cl(x_M)}{ l_p} \right) 
-  \frac{E}{2 \pi}  \log \left( \frac{ x_\cl(p_M)}{l_x}  \right) 
\nonumber \\
& + &  \int_{l_x}^{x_M} \frac{dx}{2 \pi}  \;x \frac{d p_\cl(x)}{d x} +
\int_{l_p}^{p_M} \frac{dp}{2 \pi}  \;p \frac{d  x_\cl(p)}{d p}  
\nonumber    
\earray
\no
The BK conditions (\ref{s12}) of course reproduce
eq. (\ref{s4}). More general 
boundary functions induce a fluctuation
term in the counting formula of a form which recalls
eq.(\ref{s6}). Let us denote this term as
\barray 
 n_{\rm fl} (E) &  = &  
 -  \frac{E}{2 \pi}  \log \left( \frac{ p_\cl(x_M)}{ l_p} \right) 
-  \frac{E}{2 \pi}  \log \left( \frac{ x_\cl(p_M)}{l_x}  \right) 
\label{s20} \\
& + &  \int_{l_x}^{x_M} \frac{dx}{2 \pi}  \;x \frac{d p_\cl(x)}{d x} +
\int_{l_p}^{p_M} \frac{dp}{2 \pi}  \;p \frac{d  x_\cl(p)}{d p}   
\nonumber    
\earray
\no
so that
\beq
\CN(E) =  \CN_{BK}(E) + n_\fl(E) 
\label{s20-1}
\eeq
\no
Taking the derivative of (\ref{s20}) with 
respect to $E$,  and using eqs.(\ref{s17}) 
one gets 
\beq
\frac{ d  n_{\rm fl} (E)}{dE} 
= -  \frac{1}{2 \pi}  \log \left( \frac{ p_\cl(x_M)}{ l_p} \right) 
-  \frac{1}{2 \pi}  \log \left( \frac{ x_\cl(p_M)}{l_x}  \right)
\label{s21}
\eeq

\no
which implies that the boundary functions
are related to the  fluctuation part 
of the density of states. A further simplification
is achieved imposing the $x p$  symmetry (\ref{s15})

\beq
\frac{ p_\cl(x_M)}{ l_p} = \frac{ x_\cl(p_M)}{l_x}, 
\;\; \frac{p_{M}}{l_p} = \frac{x_M}{l_x}   
\label{s22}
\eeq

\no
which leads to

\beq
\frac{ d  n_{\rm fl} (E)}{dE} 
= -  \frac{1}{\pi}  \log \left( \frac{ p_\cl(x_M)}{ l_p} \right) 
= -  \frac{1}{\pi}  \log \left( \frac{ x_\cl(p_M)}{l_x}  \right)
\label{s23}
\eeq

\no
Hence,  $x p$-symmetric boundary functions  
$p_\cl(x_M)$ and  $x_\cl(p_M)$
are completely fixed by the density of the fluctuations.  
To find $p_\cl(x)$, one combines
(\ref{s23}) and (\ref{s17})

\beq
p_\cl(x_M) = l_p \; e^{ - \pi n'_{\rm fl}(E)} = \frac{E}{x_M}, 
\qquad n'_{\rm fl}(E) = \frac{d n_{\rm fl}(E)}{dE} 
\label{s24}
\eeq

\no
which gives  $x_M$ as a function of $E$

\beq
x_M  = \frac{E}{l_p}  \; e^{  \pi n'_{\rm fl}(E)} 
\label{s25}
\eeq

\no
If  $n_{\rm fl}(E)=0$ ,  the latter equations 
reproduce the BK boundary conditions (\ref{s13}).
Eq.(\ref{s25}) gives $x_M$ as a function of $E$
and it is monotonically increasing provided

\beq
\frac{d x_M(E)}{dE} > 0  \Longrightarrow
1 + \pi E \frac{d^2 n_{\rm fl}(E)}{dE^2}  > 0  
\label{s26}
\eeq

\no
Under this condition we can expressed $E$ as a
function of $x_M$ and  replaced it in 
(\ref{s24}), obtaining the boundary function 
$p_{x_M} = E(x_M)/x_M$. In this case the
inverse problem of finding a Hamiltonian
given the spectrum has a unique solution
at the semiclassical level. If the fluctuations
are strong enough at some energies, then condition
(\ref{s26}) could  be violated implying that   
$E = E(x)$ as well as $p_\cl(x)$ will 
be  multivalued functions. This gives rise to 
a manifold of boundary functions, each
one having discontinuities at some values
of $x$.

\section{From classical to quantum}

In this section we shall give a quantum version of the 
semiclassical
results obtained above. The starting point is the
quantization of the classical hamiltonian $H_0^\cl= xp$.
Let us consider the usual normal ordered expression
\beq
H_0 = \frac{1}{2} ( x p + p x) = - i \left(
x \frac{d}{dx} + \frac{1}{2} \right) 
\label{t1}
\eeq
\no
where $p = - id/dx$. 
In references \cite{Sierra2,Twamley} it was shown that $H_0$
becomes a self-adjoint operator in two cases
where the domain of the $x$ variable are choosen as: 
1)  $ 0 < x < \infty $ or 
2) $a < x < b$ with $a$ and $b$ finite. 
For the purposes of this paper we shall 
confine to the case 1. Case 2 was discussed at length
in \cite{Sierra2}. Since $x > 0$ one can write
(\ref{t1}) as
\beq
H_0 = x^{1/2} \;p \;  x^{1/2}, \qquad x >0 
\label{t2}
\eeq
\no
The exact eigenfunctions of (\ref{t2}) are given by
\beq
\phi_E(x) = \frac{1}{\sqrt{ 2 \pi}} \frac{1}{x^{1/2 - i E}}, \qquad
E \in \Rmath 
\label{t3}
\eeq
\no
where the eigenenergies $E$ belong to the real line. 
The normalization of (\ref{t3}) is the appropiate one
for a continuum spectra,
\beq
\langle \phi_E| \phi_{E'} \rangle  = 
 \int_0^\infty dx \;  \phi_E^*(x)  \phi_{E'}(x) =
\delta(E - E'). 
\label{t4}
\eeq
\no
The quantum Hamiltonian associated to the
semiclassical approach is 
\beq
H = H_0 + i \left( | \psi_a \rangle \langle \psi_b |
-  | \psi_b \rangle \langle \psi_a |
\right)
\label{t5}
\eeq
\no
where $\psi_a$ and $\psi_b$ are two wave functions
associated to the boundary functions $p_\cl(x)$
and $x_\cl(p)$, respectively, i.e.
\beq
\psi_a \leftrightarrow p_\cl(x), \qquad
\psi_b  \leftrightarrow x_\cl(p) 
\label{t6}
\eeq
\no
We shall choose real functions $\psi_a(x)$
and $\psi_b(x)$ so that $H$ is an hermitean
and antisymmetric operator, which implies
that the eigenvalues appear in pairs ${E, -E}$. 
The interaction term in (\ref{t5}) 
can be justified by the following heuristic 
argument. Let us consider a particle
which at $t=0$ belong  to the classical  allowed region, 
i.e. $x_0 > x_\cl(p_0)  $ and $p_0 > p_\cl(x_0)$.
According to the classical evolution (\ref{s2}),
the position $x(t)$ increases while the momenta
$p(t)$ decreases, i.e.  
 \beq
{\rm Classical} \; {\rm evolution:} (x_0, p_0) 
{\longrightarrow} (e^t x_0, e^{-t} p_0)  
\label{t7}
\eeq
\no 
until a time $t_M$ where the particle
hits the $p_\cl$-boundary. 
\begin{figure}[t!]
\begin{center}
\includegraphics[height= 7.0 cm,angle= 0]{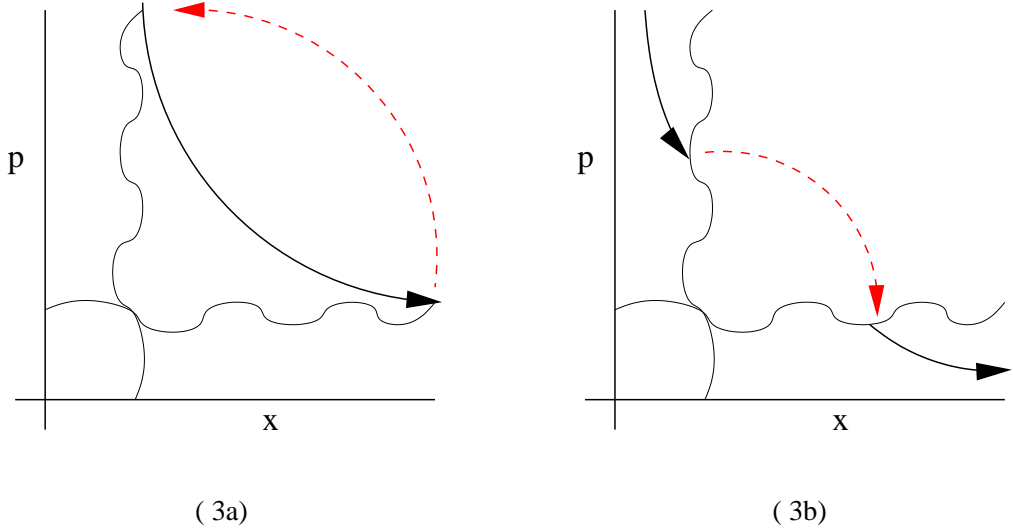}
\end{center}
\caption{
Graphical representation of the classical transport  operation in the 
phase space of the $xp$ model given 
in eqs.(\ref{t9}) and (\ref{t15}).  
}
\label{teletransport}
\end{figure}
 \beq
 (e^{t_M} x_0, e^{-t_M} p_0) = (x_M, p_\cl(x_M))   
\label{t8}
\eeq
\no 
The semiclassical approach
suggests to transport this particle from the $p_\cl$-boundary
to a point in the  $x_\cl$-boundary, (see fig. \ref{teletransport})
\beq
{\rm Classical} \; {\rm transport:}  (x_M, p_\cl(x_M)) \rightarrow
(x_\cl(p_M), p_M), 
\label{t9}
\eeq
\no
while preserving the energy,
\beq
E = x_0 \; p_0 = x_M \;  p_\cl(x_M) = x_\cl(p_M) \;  p_M    
\label{t10}
\eeq
\no
Equation (\ref{t10}) coincides with (\ref{s17})
if we choose $(x_0, p_0) = (x_I, p_I)$. 
The transported particle at the $x_\cl$- boundary
continues its classical evolution returning 
to the initial point $(x_0, p_0)$ after a time
\beq
\tau_E = 
\frac{1}{E} \log \frac{ x_M}{x_\cl (p_M) } =  
\frac{1}{E} \log \frac{p_M}{ p_\cl (x_M) }
\label{t11}
\eeq
\no
This is also the period of the classical
trayectory which has become a closed orbit thanks to the 
transport operation (\ref{t9}). The semiclassical
calculation of the previous section  measures 
classical action associated to this periodic
orbit. 
At the quantum level the free evolution 
of a state $\psi$ is given by the
unitary transformation 
 \beq
{\rm Quantum} \; {\rm evolution:}  |\psi(0) \rangle
{\longrightarrow} |\psi(t) \rangle = e^{- i t H_0} |\psi(0) \rangle 
\label{t12}
\eeq
\no 
The operator that performs the transport (\ref{t9}) 
is given by one of the interacting terms in the Hamiltonian
(\ref{t5}), 
\beq
{\rm Quantum} \;  {\rm transport:}  
|\psi \rangle \rightarrow -i |\psi_b\rangle \langle \psi_a| \psi \rangle 
\label{t13}
\eeq
\no
which consists in the proyection of the state $\psi$ into 
the quantum state $\psi_a$, yielding the state $\psi_b$ as a result.  
The hermiticity of the Hamiltonian $H$ implies the
existence of the inverse of the process (\ref{t13}), i.e. 
\beq
{\rm Inverse} \; {\rm quantum} \;  {\rm transport:}  
|\psi \rangle \rightarrow -i |\psi_a\rangle \langle \psi_b| \psi \rangle 
\label{t14}
\eeq
\no
whose classical analogue is (see fig. \ref{teletransport}b), 
\beq
{\rm Classical} \; {\rm Inverse} \; {\rm transport:} \; (x_\cl(p_M), p_M),
 \rightarrow  (x_M, p_\cl(x_M))
\label{t15}
\eeq
\no
What is the physical meaning of this process?
Let us take for a while a particle in the classical forbbiden region where
$x_0 < x_\cl(p_0) $ but  $p_0 >  p_\cl(x_0)$. This particle will evolve
freely according to eqs.(\ref{t7}), until a time $t_M$
where it hits the $x_\cl$-boundary, i.e. 
 \beq
 (e^{t_M} x_0, e^{-t_M} p_0) = (x_\cl(p_M), p_M))   
\label{t16}
\eeq
\no
Then one can apply the inverse transport (\ref{t15})
which carries the particle to the $p_\cl$-boundary where
it continues its free and unbounded 
evolution : $x \rightarrow \infty$
and $p \rightarrow 0$. The phase space area 
traced by this trayectory is infinite 
which implies that
the number of these kind of 
semiclassical states is infinite forming therefore 
a continuum.

In summary, the transport operations  
between the two boundaries leads classically
to closed periodic trayectories in the allowed
phase space and to open trayectories in the forbbiden
region. Semiclassically the closed periodic
trayectories give rise to bound states while the open
ones form a continuum. This is scenario that comes
out from the solution of the quantum model, as we show 
in  the next section. 

The existence of a semiclassical continuum 
in the $x p$ model was proposed
by Connes in  reference \cite{Connes}. 
Instead of the boundary conditions 
set by $l_x$ and $l_p$, Connes 
restricts the phase space of the model to be 
$|x| < \Lambda$, $|p| < \Lambda$,  where
$\Lambda$ is a cutoff which is sent to infinite
at the end of the calculation. The number of semiclassical
states is given now by 
\beq
\CN_C(E)   =   \frac{E}{ \pi} \log \Lambda
- \frac{E}{2 \pi} \left( \log \frac{E}{2 \pi} -1 \right)
\label{t17}
\eeq
\no
where the first term leads, in the limit $\Lambda \rightarrow
\infty$, to a continuum while the second term
coincides with minus the average position
of the Riemann zeros (\ref{s4}).
A possible interpretation of these result
is that the Riemann zeros, are missing spectral
lines in a continuum, which is in apparent contradiction 
with the Berry-Keating interpretation of the zeros
as bound states. As we shall show below
both interpretations can be  reconciled at the quantum
level where the Riemann zeros appear as discrete
spectra embbeded in a continuum of states.

\section{Exact solution of the Schroedinger equation}

In this section we shall find explicitely the
eigenstates and the eigenergies of the Hamiltonian (\ref{t5})
for generic states $\psi_a$ and $\psi_b$. 
The method used is similar to the one employed in
reference \cite{Sierra2}, where instead of the Hamiltonian
$xp$ we added an interaction to $1/xp$. 
The Schroedinger equation for an eigenstate
$\psi_E(x)$ with energy $E$ is given by 
\beq
 - i \left( x \frac{d}{dx} + \frac{1}{2} \right) \psi_E(x) 
+ i  \left( \psi_a(x) \langle \psi_b| \psi_E \rangle
- \psi_b(x) \langle \psi_a| \psi_E \rangle =  E \psi_E(x)
\right) 
\label{u1} 
\eeq
\no
Let us introduce the variable $q$
\beq
q= \log x, \qquad q \in \Rmath 
\label{u2}
\eeq
\no
and the overlap integrals
\barray 
A & = &  \langle \psi_a| \psi_E \rangle = 
\int_0^\infty dx \; \psi_a(x) \psi_E(x)  
\label{u3} \\
B & = &  \langle \psi_b| \psi_E \rangle = 
\int_0^\infty dx \; \psi_b(x) \psi_E(x)  
\nonumber 
\earray 
\no
which depend on $E$. Using these definitions
eq.(\ref{u1}) becomes
   
\beq
- i \left( \frac{d}{dq} + \frac{1}{2} \right) 
\psi_E(q) + i ( B \psi_a(q) - A \psi_b(q) ) = E \psi_E(q) 
\label{u4}
\eeq
\no
The general solution of this equation is given by
\beq
\psi_E(q)    =   
    e^{- (1/2 - i E) q } \left[ 
C_0 + \int_{-\infty}^q  dq' \;  e^{(1/2 - i E) q'}
( B \psi_a(q') - A \psi_b(q'))  
\right] 
\label{u5} 
\eeq
\no
where $C_0$ is an  integration constant. 
It is convenient to define the  functions
\barray 
a(q)  & = & e^{q/2} \psi_a(q), \qquad
\psi_a(x) = \frac{a(x)}{\sqrt{x}} 
\label{u6} \\
b(q)  & = & e^{q/2} \psi_b(q), \qquad
\psi_b(x) = \frac{b(x)}{\sqrt{x}} 
\nonumber 
\earray 
\no 
so that 
\beq
\psi_E(q)   =
    e^{- (1/2 - i E) q }
\left[ 
C_0 + \int_{-\infty}^q  dq' \;  e^{- i E q'}
( B \; a(q') - A \;  b(q'))  
\right] 
\label{u7}
\eeq
\no
An alternative way to express (\ref{u7})  is
\beq
\psi_E(q)   = 
    e^{- (1/2 - i E) q }
\left[ 
C_\infty - \int_q^\infty dq' \;  e^{- i E q'}
( B \; a(q') - A \;  b(q'))  
\right] 
\label{u8}
\eeq
\no
where $C_\infty$ is related to $C_0$ by
\beq
C_\infty = C_0 + B \; \ha(-E) - A \; \hb(-E) 
\label{u9}
\eeq
\no
where 
\beq
\hf(E) = \int_{- \infty}^\infty dq \; e^{i E q} f(q), 
\qquad f= a, b 
\label{u10}
\eeq
\no
We shall assume that $a(q)$ and $b(q)$
satisfy 
\barray 
& \lim_{q \rightarrow - \infty} \int_{-\infty}^q d q'\; e^{- i E q'}
f(q') = 0, \quad f = a, b  & 
\label{u11} \\
& 
 \lim_{q \rightarrow  \infty} \int_q^{\infty} d q'\; e^{- i E q'}
f(q') = 0, \quad f = a, b
& 
\nonumber 
\earray 
\no
which implies that the asymptotic behaviour of $\psi_E(x)$
is dominated by $C_0, C_\infty$, i.e.
\beq
\lim_{x \rightarrow 0} \psi_E(x) = \frac{C_0}{x^{1/2 - i E}}, 
\quad
\lim_{x \rightarrow \infty} \psi_E(x) = \frac{C_\infty}{x^{1/2 - i E}}, 
\label{u12}
\eeq
\no
Plugging (\ref{u7}) into (\ref{u3}) yields
the relation between the constants $A,B,C_0$,
\beq
\left( 
\begin{array}{cc}
1 + S_{a,b} & - S_{a,a} \\
S_{b,b} & 1 - S_{b,a} \end{array}
\right) 
\left( 
\begin{array}{c}
A \\
B
\end{array}
\right) 
= C_0 
\left( 
\begin{array}{c}
\ha(E) \\
\hb(E) 
\end{array}
\right) 
\label{u13}
\eeq
\no
where the functions $S_{f,g}(E)$ with $f,g = a,b$ 
are defined by \cite{s-function}
\beq
S_{f,g}(E) =
\int_{- \infty}^\infty dq \;  e^{i E q} \;
f(q) \int_{-\infty}^q dq' \;  e^{-i E q'} \, g(q') 
\label{u14}
\eeq
\no
Similarly, introducing (\ref{u8}) into
(\ref{u3}) yields
\beq
\left( 
\begin{array}{cc}
1 - \tS_{a,b} &  \tS_{a,a} \\
- \tS_{b,b} & 1 + \tS_{b,a} \end{array}
\right) 
\left( 
\begin{array}{c}
A \\
B
\end{array}
\right) 
= C_\infty 
\left( 
\begin{array}{c}
\ha(E) \\
\hb(E) 
\end{array}
\right) 
\label{u15}
\eeq
\no
where 
\beq
\tS_{f,g}(E) =
\int_{- \infty}^\infty dq \;  e^{i E q} \;
f(q) \int_q^{\infty} dq' \;  e^{-i E q'} \, g(q') 
\label{u16}
\eeq
\no
This function   is related to $S_{f,g}$ in two ways, 
\barray 
\tS_{f,g}(E) &  = & - S_{f,g}(E) + \hf(E) \;  \hg(-E) 
\label{u18} \\
\tS_{f,g}(E) &  = & S_{g,f}(-E)
\label{u19}
\earray 
\no
To derive these equations one makes a change of order in the integration.
Combining 
(\ref{u18}) and (\ref{u19}) one obtains the {\em shuffle} relation
\beq
S_{f,g}(E) + S_{g,f}(-E) 
= \hf(E) \;  \hg(-E)
\label{u20}
\eeq
\no
The terminology is borrowed from the theory
of multiple zeta functions where there is a similar
relation between
the two variable Euler-Zagier zeta function 
$\zeta(s_1, s_2)$, and the Riemann zeta
function $\zeta(s)$ \cite{euler-zagier-1,euler-zagier-2}. 

The solutions of the eqs.(\ref{u13}) and (\ref{u15})
depend on the determinant of the associated $2 \times 2$
matrices  given by
\barray
\F(E)& = & 1 + S_{a,b} - S_{b,a} + S_{a,a} S_{b,b} - S_{a,b} S_{b,a}
\label{u21} \\
 \tF(E)& = & 1 - \tS_{a,b} + \tS_{b,a} + \tS_{a,a} \tS_{b,b} - 
\tS_{a,b} \tS_{b,a}
\label{u22} 
\nonumber 
\earray  
\no
which are related by (\ref{u19})
\beq
\tF(E) = \F(-E) 
\label{u23}
\eeq
\no
Moreover, since
$a(x)$ and $b(x)$ are real functions one has
\beq
S_{f,g}^*(E) =S_{f,g}(-E^*)    
\label{u24}
\eeq
\no
which in turn implies
\beq
\F^*(E) = \F(-E^*) 
\label{u25}
\eeq
\no
After these observations we can return
to the solution of (\ref{u13}) and (\ref{u15}).
We shall distinguish two cases: 1) $\F(E) \neq 0$
and 2) $\F(E) = 0$, where $E$ is real since it is 
an eigenvalue of the Hamiltonian
(\ref{t5}).

{\bf Case 1:}  $\F(E) \neq 0$

 Eq.(\ref{u25}) implies  that   $\F(-E) \neq 0$
and therefore $A$ and $B$ can be expressed
in two different ways, 
\barray
A & = & \frac{C_0}{\F(E)} 
\left[ ( 1 - S_{b,a}) \;  \ha(E) + S_{a,a} \;  \hb(E) \right]
\nonumber \\
& = &   \frac{C_\infty}{\F(-E)} 
\left[ (1 + \tS_{b,a}) \;  \ha(E) - \tS_{a,a} \; \hb(E) \right], 
\label{u26} \\
B & = & \frac{C_0}{\F(E)} 
\left[ - S_{b,b} \; \ha(E) + ( 1 + S_{a,b}) \;  \hb(E) 
 \right] 
\nonumber 
\\
 & = & \frac{C_\infty}{\F(-E)} 
\left[ \tS_{b,b} \; \ha(E) + ( 1- \tS_{a,b}) \; \hb(E) 
\right]
\earray 
\no
Now using eq.(\ref{u18}),  these eqs. reduce to
\beq
\frac{C_0}{C_\infty} = \frac{\F(E)}{\F(-E)} 
\label{u27}
\eeq
\no
which by eq.(\ref{u25}) is a pure phase for $E$ real. 
Hence, up to an overall factor, the integration
constants for this solution can be choosen as
\barray
C_0 & = & \F(E) \nonumber \\
C_\infty & = & \F(-E) 
\label{u28} \\
A & = & 
 ( 1 - S_{b,a}) \;  \ha(E) + S_{a,a} \;  \hb(E)
\nonumber \\
B & = &  - S_{b,b} \; \ha(E) + ( 1 + S_{a,b}) \;  \hb(E). 
\nonumber 
\earray 
\no
Since the constants $C_0, C_\infty$ do not vanish, 
the wave function is non normalizable near the origin
and  infinity (recall eq. (\ref{u12})) and therefore
they correspond to scattering states. Of course
they will be normalizable in the distributional sense.

{\bf Case 2:} $\F(E) = 0$.

The integration constants
can be choosen as
\barray
C_0 & = & 0 \nonumber \\
C_\infty & = & 0 
\label{u29} \\
A & = &  S_{a,a} 
\nonumber \\
B & = &  ( 1 + S_{a,b}) 
\nonumber 
\earray 
\no
which solves eqs. (\ref{u13}) and (\ref{u15}). 
Since $C_0=  C_\infty = 0 $, the leading term
of the behaviour of $\psi_E(x)$ vanish
near the origin and  infinity and under
appropiate conditions on  $\psi_{a,b}$,  the state
$\psi_E$ 
will be normalizable corresponding to a 
bound state. In the appendix we compute the norm
of these states. 

Hence the generic spectrum of the Hamiltonian
(\ref{t5}) consist of a continuum covering the
whole real line with, eventually, some isolated bound
states embedded in it, whenever $\CF(E) = 0$.
This structure also arises in the Hamiltonian
studied in reference \cite{Sierra2}. The function
$\CF(E)$ plays the role of the Jost function
since its zeros gives the position of the bound
states and its phase gives the scattering phase shift
according to eq.(\ref{u27}). 

Before we continue with the general formalism  it
is worth to study a simple case which illustrates
the results obtained so far.

\begin{figure}[t!]
\begin{center}
\includegraphics[height= 2.5 cm,angle= 0]{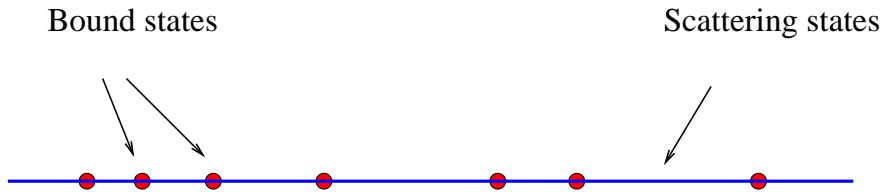}
\end{center}
\caption{
Pictorial representation of the spectrum of the model.
The bound states are the points where $\F(E) = 0$, which
are embedded in a continuum of scattering states. 
}
\label{bound-states}
\end{figure}

\subsection*{An example: a quantum trap} 

Let us start with the classical version
of a trap where a particle is restricted to the
region $x_b < x < x_a$.
The semiclassical number of states is given 
by the area formula (\ref{s3}), 
\beq
n = \frac{A}{2 \pi} = \int_{x_b}^{x_a} \frac{dx}{2 \pi} 
\; \frac{E}{x} = \frac{E}{2 \pi} \log \frac{x_a}{x_b}
\label{u30}
\eeq
\no
which yields the eigenenergies
\beq
E_n = \frac{2 \pi \; n}{\log(x_a/x_b)},\qquad n \in \N.
\label{u31}
\eeq
\no
The quantum version of this model is realized by 
two boundary states $\psi_{a,b}(x)$ 
proportional to delta functions, i.e. 
\beq
\psi_a(x) = a_0 \; x_a^{1/2}
 \delta(x - x_a), \; 
\psi_b(x) = b_0 \; x_b^{1/2}
 \delta(x - x_b).
\label{u32-1}
\eeq
\no 
The associated potentials $a(q), b(q)$ are 
\barray 
& a(q) = a_0 \delta(q - q_a), \; b(q) = b_0 \delta(q - q_b), 
&  \label{u32} \\
& q_{a} = \log x_{a}, \; q_{b} = \log x_{b}  &
\nonumber 
\earray 
\no
The various quantities defined above are readily 
computed obtaining 
\barray
\ha & = & a_0 \; e^{i E q_a}, \qquad 
\hb  =  b_0 \; e^{i E q_b} \nonumber 
\\
S_{a,a} & = & \frac{a_0^2}{2},  \qquad S_{b,b} = \frac{b_0^2}{2} 
\label{u33} \\
S_{a,b} &  = &  a_0 \; b_0 \; e^{i E q_{a,b}}  
\qquad
S_{b,a} = 0 
\nonumber 
\earray

\begin{figure}[t!]
\begin{center}
\includegraphics[height= 7.0 cm,angle= 0]{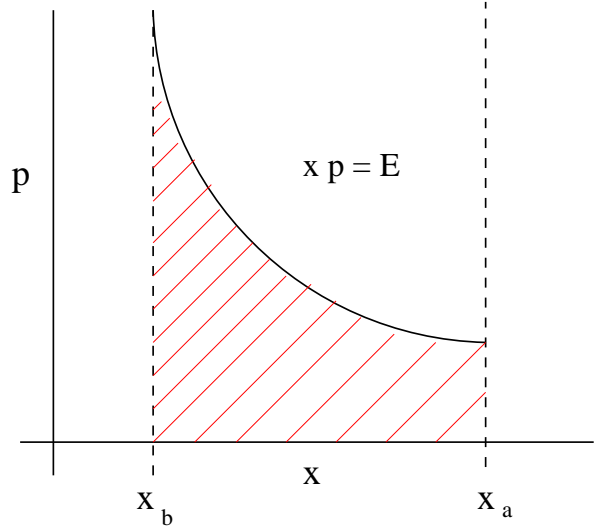}
\end{center}
\caption{
Semiclassical picture of the model represented by the potential
(\ref{u32}). 
}
\label{trap}
\end{figure}
\no
where  $q_{a,b} = q_a - q_b = \log(x_a/ x_b)$. Plugging these eqs. into
(\ref{u21}) yields 
\beq
\CF(E) = 1 +  \left(  \frac{a_0 b_0}{2} \right)^2
+ a_0 b_0 \;  e^{i E \;  q_{a,b}} 
\label{u34}
\eeq
\no
For generic values of $a_0, b_0$, 
the Jost function (\ref{u34}) never
vanishes obtaining a spectrum which 
is continuous. However, $\CF(E)$
vanishes provided the following condition holds
\beq
\ep \equiv \frac{a_0 b_0}{2} = \pm 1 \Longrightarrow
\CF(E) = 2 ( 1 + \ep \;  e^{i E \; q_{a,b}}) 
\label{u35}
\eeq
\no
in which cases the spectrum contains bound states embbeded
in the continuum with energies
\barray
{\rm If} \; \ep & = &  1    \Longrightarrow   
 E_n = \frac{2 \pi ( n + 1/2)}{ q_{a,b} } \;\; \; n \in \N 
\label{u36} \\
{\rm If} \; \ep & = &  - 1    \Longrightarrow   
 E_n = \frac{2 \pi n}{ q_{a,b} } \;\;\;  n \in \N 
\nonumber 
\earray 
\no
that agree with the semiclassical energies
(\ref{u31}) for $n >> 1$. 
The unnormalized wave function of the bound states, 
i.e. $\CF(E) = 0$, 
can be computed from eq. (\ref{u7})
\beq
\psi_{E}(x) = 
 \frac{1}{x^{1/2 - i E}} \times 
\left\{ 
\begin{array}{ll} 
1, & x_b < x < x_a \\
0, & x < x_b \; {\rm or} \; x > x_a 
\end{array}
\right. 
\label{u37}
\eeq
\no
which shows that they are confined
to the region $(x_b, x_a)$. The wave functions
when $\CF(E) \neq 0$ can be similarly found obtaining
\beq
\psi_{E}(x) = \frac{1}{x^{1/2 - i E}} \times  
\left\{ 
\begin{array}{ll} 
\CF(E), & 0 < x < x_b  \\
1 -  \left( \frac{a_0 b_0}{2} \right)^2, & x_b < x < x_a \\
\CF(-E), & x_a < x < \infty \\
\end{array}
\right. 
\label{u38}
\eeq
\no
Hence if (\ref{u35}) holds, these wave functions
vanishes in the region $(x_b, x_a)$ which
contains the trapped particles (\ref{u37}). 
In this example the mechanism responsible 
for the existence of bound states is the transport of the 
particles from the position $x_a$ to the position $x_b$. 
At the quantum level
the confinement requires the fine tuning  of the couplings
(see  eq. (\ref{u35})), which  introduces periodic 
or antiperiodic  boundary conditions 
depending on the sign of $\ep$. When $|\ep| \neq 1 $ 
the particle can scape the trap and the bound states 
become resonances.

\section{Analyticity properties of $\CF(E)$}

As in ordinary Quantum Mechanics, 
the Jost function $\CF(E)$ satisfy 
certain analyticity properties reflecting the causal
structure of the dynamics.
In our case these properties follows from those
of the function $S_{f,g}$ (eq. (\ref{u14}))  
and the definition (\ref{u21}). 

Indeed, let us express $S_{f,g}(E)$
in terms of the Fourier transforms
of the functions $f,g$. First we replace
$g(q)$ by its inverse Fourier transform
\beq
g(q') = \int_{- \infty}^\infty \frac{dE'}{2 \pi} e^{i E q'} \; \hg(-E')
\label{a1}
\eeq
\no
back into eq.(\ref{u14}), obtaining 
\beq
S_{f,g}(E)   =  \int_{- \infty}^\infty \frac{dE'}{2 \pi} \hg(-E')
 \int_{- \infty}^\infty dq \; e^{i E q} f(q) 
 \int_{- \infty}^q  dq' \; e^{i (E' - E) q'}.
\label{a2} \eeq
\no
The last integral is given by the distribution
\beq
\int_{- \infty}^q  dq' \; e^{i (E' - E) q'} = 
 e^{i q (E' - E)} \left[ \pi \delta(E'- E) +
\frac{1}{i} P \; \frac{1}{E' - E} \right] 
\label{a3}
\eeq
\no
where $P$ denotes the Cauchy principal part. 
Plugging (\ref{a3}) into (\ref{a2}) and using the Fourier
transform of $f$ gives,
\beq
S_{f,g}(E) = \frac{1}{2} \left[
\hf(E) \; \hg(-E) + P \int_{-\infty}^\infty
\frac{dE'}{\pi i} \frac{ \hf(E') \; \hg(-E')}{E' - E} 
\right]
\label{a4}
\eeq
\no
Alternatively,  one can write (\ref{a4}) as
\beq
S_{f,g}(E) =  \int_{-\infty}^\infty
\frac{dE'}{2 \pi i} \frac{ \hf(E') \; \hg(-E')}{E' - E - i \ep} 
\label{a5}
\eeq
\no
with $\ep > 0$ an infinitesimal. 
Eq. (\ref{a5})  shows that the poles of $S_{f,g}(E)$ are located
in the lower half of the complex energy plane. Thus for 
well behave functions $\hf, \hg$, the function
$S_{f,g}(E)$ will be  analytic in the
complex upper-half plane.  These 
properties also apply to $\CF(E)$ which 
is  the product of $S_{f,g}$ functions with 
$f,g = a,b$. 
\no
Another important property of the Jost function $\CF(E)$
is that its zeros lie either on the real axis or below it, i.e.
\beq
{\rm If} \; \CF(E) = 0 \Longrightarrow
\Im \; E \leq 0  
\label{a6}
\eeq
\no
The proof of this equation is similar to the
one done in reference \cite{Sierra2}, being convenient
to regularize the interval $x \in (0, \infty )$
as $(N^{-1}, N)$ with $N \rightarrow \infty$. 

In the appendix we use the results obtained in this section
to compute the norm of the eigenstates. 

\section{The quantum version of the
Berry-Keating  model}

Let us consider the BK constraints $x > l_x$ and $|p| > l_p$. 
It is rather natural to associate 
constraint $x > l_x$ with the wave function
\beq
\psi_b(x) = b_0 \; l_x^{1/2} \;  \delta(x- l_x)
\label{r1}
\eeq
\no
which is localized at the boundary $x = l_x$. 
The factor $\l_x^{1/2}$ gives the correct
dimensionality  to $\psi_b(x)$,  with $b_0$
a dimensionaless parameter. 
On the other hand the constraint $|p| > l_p$
admits  two possible quantum versions, 
\beq
\left\{ 
\begin{array}{c}
\psi_a^+ (x) \\
\psi_a^-(x) \\
\end{array}
\right. 
 = 2 a_0 \left( \frac{l_p}{2 \pi} \right)^{1/2}  \times 
\left\{ 
\begin{array}{c}
\cos( l_p x) \\
\sin( l_p x) 
\end{array}
\right. 
\label{r2}
\eeq
\no 
Due to the fact that $\psi_a$ has to be real, 
one cannot choose a pure plane wave  $e^{i l_p x}$. 
The boundary wave functions (\ref{r1}) and
(\ref{r2}) are  the cosine and sine Fourier
transform of each other, namely
\beq
\left\{ 
\begin{array}{c}
\psi_a^+ (x) \\
\psi_a^-(x) \\
\end{array}
\right. 
= \frac{2 a_0}{b_0}  \left( \frac{l_p}{2 \pi l_x} \right)^{1/2}
\int_0^\infty dy
\; \psi_b(y) \times 
\left\{ 
\begin{array}{c}
\cos ( l_p x y/l_x) \\
 \sin ( l_p x y/l_x) \\
\end{array}
\right.  
\label{r3}
\eeq
\no
Indeed, extending the domain of $\psi_b(x)$ according to the
parity of $\psi_a^\eta \; (\eta = \pm) $ one gets
\beq
 \psi_b(-x) = \eta \psi_b (x)  \rightarrow 
 \psi_a^\eta (x) = \frac{a_0}{b_0}   
\left( \frac{l_p}{2 \pi l_x} \right)^{1/2}
e^{i \frac{\pi}{4} (\eta -1)}
\hat{\psi}_b \left( \frac{ l_p x}{l_x} \right)
\label{r4}
\eeq
\no
which are the quantum analogue of 
the classical equations (\ref{s15}).
Later on, we shall consider more
general wave functions $\psi_{a,b}$
to account for the fluctuations 
in the Riemann formula, imposing again eq.(\ref{r3}).
The relation (\ref{r3}) between $\psi_a^{\pm}$
and $\psi_b$ must imply a close link 
between their Mellin transforms  
$\ha_\pm(E)$ and $\hb(E)$. To derive it,  
let us write
\beq
\ha_\pm(E)  = 
 \int_0^\infty x^{-1/2 + i E} \;  \psi_a^\pm (x) 
 =  \frac{2 a_0}{b_0}   \left( \frac{l_p}{2 \pi l_x} \right)^{1/2}
\int_0^\infty dx \; x^{-1/2 + i E} 
\int_0^\infty dy \; \psi_b(y) 
\times 
\left\{
\begin{array}{c}
\cos( l_p x y/l_x) \\
\sin( l_p x y/l_x) \\
\end{array}
\right. 
\label{r5}
\eeq
\no
The basic integrals one needs  are
\beq
\int_0^\infty dx \; x^{-\frac{1}{2} + i E} 
\times 
\left\{
\begin{array}{c}
\cos( p x) \\
\sin(p x) \\
\end{array}
\right. 
= \frac{1}{2} \left( \frac{2 \pi}{|p|} \right)^{\frac{1}{2} + i E}
\times 
\left\{
\begin{array}{c}
e^{2 i \theta_+(E)}  \\
e^{2 i  \theta_-(E)} \\
\end{array}
\right. 
\label{r6}
\eeq
\no
where
\beq
e^{2 i \theta_{\pm}(E)} = 
\left\{
\begin{array}{lll}
\pi^{- i E} \frac{\Gamma(1/4 + i E/2)}{ \Gamma(1/4 - i E/2)},
& & \eta = + \\
\pi^{- i E} \frac{\Gamma(3/4 + i E/2)}{\Gamma(3/4 - i E/2)},
& & \eta = - \\
\end{array}
\right. 
\label{r7}
\eeq
\no
The function $\theta_+(E)$ coincides
with the phase of the Riemann zeta function
(\ref{s7}), and more generally of the even Dirichlet L-functions, 
 while  $\theta_-(E)$
is the phase factor of the odd  Dirichlet L-functions. 
 These phases 
appear in the functional relation of
even and odd $L$ functions, and they
arise in our context from the two  possible relations
between the boundary functions $\psi_a^\pm$ 
and $\psi_b$. Plugging eq.(\ref{r6})
into (\ref{r5}) yields
\beq
\ha_\pm(E)  = 
  \frac{a_0}{b_0} 
\left( \frac{ 2 \pi l_x}{l_p} \right)^{ i E} 
e^{2 i \theta_\pm(E)}  
\int_0^\infty dy \; \psi_b(y) y^{- \frac{1}{2} - i E}
\label{r8} 
\eeq
\no
where the integral is nothing but $\hb(-E)$, thus
\beq
\ha_\pm(E)  = 
  \frac{a_0}{b_0} 
\left( \frac{ 2 \pi l_x}{l_p} \right)^{ i E} 
e^{2 i \theta_\pm(E)}   \; \hb(-E) 
\label{r9} 
\eeq
\no
This important equation reflects the relation (\ref{r3})
which in turn is the quantum version of the $x p$ symmetry 
between boundaries. In the BK case, the Mellin transforms of the associated  wave functions
(\ref{r1}) and (\ref{r2})  are 
\beq
\ha_\pm(E)  = a_0
\left( \frac{ 2 \pi}{l_p} \right)^{ i E} 
e^{2 i \theta_\pm(E)},\qquad 
\hb(E) = b_0 \; l_x^{ i E} 
\label{r10} 
\eeq
\no
which are pure phases, up
to overall constants. The $S_{f,g}$ functions
can be readily computed using eq.(\ref{a4}). 
To do so,  we first consider the products
\barray
& \ha_\pm (E) \;  \ha_\pm (-E) = a_0^2, & \nonumber \\
&  \hb(E) \;  \hb(-E) = b_0^2 & 
\label{r11} \\
& \ha_\pm (E) \;  \hb(-E) = a_0 b_0 \; e^{2 i \theta_\pm(E)} 
& \nonumber \\
& \hb (E) \;  \ha_\pm(-E) = a_0 b_0 \; e^{- 2 i \theta_\pm(E)} 
& \nonumber
\earray 
\no
where we used $\l_x l_p = 2 \pi$ and that $\theta_\pm(-E) 
= - \theta_\pm(E)$. The diagonal terms of $S_{f,g}$ 
are given simply by 
\beq
S_{a_\pm, a_\pm}(E) = \frac{a_0^2}{2},
\qquad 
S_{b,b}(E) = \frac{b_0^2}{2} 
\label{r12}
\eeq
\no
since the Hilbert transform of a constant is zero, i.e.
\beq
P \int_{- \infty}^\infty \frac{dt}{\pi i} \frac{1}{t - E} = 0, \qquad
E \in \Rmath
\label{r13}
\eeq
\no
The computation of $S_{a_\pm,b}$ and $S_{b,a_\pm}$ 
uses the analytic properties of $e^{2 i \theta_\pm(E)}$.
Let us focus on the case of  $e^{2 i \theta_+(E)} = e^{2 i \theta(E)}$.
This function converges rapidly to zero as $|E| \rightarrow \infty$  in the upper half plane, 
and it  has poles at $E_n = i ( 2 n + 1/2) \;\; (n=0, 1, \dots)$
where it behaves like 
\beq
 e^{2 i \theta(E)}\sim \frac{(-1)^n \; 2 ( 2 \pi)^{2 n}}{(2 n)!} 
\frac{1}{2 n + 1/2 + i E} 
\label{r14}
\eeq
\no
We can split $e^{2 i \theta(E)}$ into the sum
\barray 
 e^{2 i \theta(E)} & = & \Omega_+(E) + \Omega_-(E)  
\label{r15} \\
\Omega_-(E)& = & \sum_{n=0}^\infty 
\frac{(-1)^n \; 2 ( 2 \pi)^{2 n}}{(2 n)!} 
\frac{1}{2 n + 1/2 + i E}  
\nonumber 
\earray 
\no
where $\Omega_+(E)$ is analytic in the upper half
plane and goes to zero at $+ i \infty$, while
$\Omega_- (E)$ has poles in the upper half plane
and behaves as $1/E$ at infinity. The function
$\Omega_-(E)$ can also be written as
\barray 
\Omega_-(E) &  =  & 2 \int_0^1 dx \; x^{- 1/2 + i E} \cos( 2 \pi x) 
\label{r16} \\
& = & \frac{4}{1 + 2 i E}\;  _1F_2 ( \frac{1}{4} + i \frac{E}{2},
\frac{1}{2},  \frac{5}{4} + i \frac{E}{2}, - \pi^2) 
\nonumber 
\earray 
\no
where $_1F_2$ is a hypergeometric function of the type $(1,2)$. 
From the analyticity properties of $\Omega_\pm$
one gets inmediately their Hilbert transform
\beq
P \int_{- \infty}^\infty \frac{dt}{\pi i} \frac{\Omega_{\pm}(t)}{t - E} 
= \pm \; \Omega_\pm(E), \qquad
E \in \Rmath
\label{r17}
\eeq
\no
Hence $S_{a_+,b} \equiv S_{a,b}$, as given by
eq.(\ref{a4}), becomes 
\barray
S_{a,b}(E) & = & \frac{a_0 b_0}{2} 
\left[ e^{2 i \theta(E)} + 
P \int_{- \infty}^\infty \frac{dt}{\pi i} \frac{ e^{2 i \theta(t)} }{t - E} 
\right]
\label{r18}
\\
& = & \frac{a_0 b_0}{2} \left[ \Omega_+(E) + \Omega_-(E)
+  \Omega_+(E) - \Omega_-(E) \right] \nonumber \\
& = &  a_0 b_0  \;  \Omega_+(E) 
\nonumber 
\earray 
\no
Similarly one finds
\beq
S_{b,a}(E) =  a_0 b_0 \;  \Omega_-(-E) 
\label{r19}
\eeq
\no
Notice that both functions are analytic in the upper
half plane. The Jost function finally reads
\barray 
\CF(E)&  =  &  1 + a_0 b_0 (\Omega_+(E) - \Omega_-(-E)) 
+ \left( \frac{a_0 b_0 }{2} \right)^2 \nonumber \\
&  & -  (a_0 b_0)^2 \Omega_+(E) \Omega_-(-E)
\label{r20} 
\earray 
\no
In the asymptotic limit $|E| >> 1$ 
\beq
\Omega_-(E) \sim \frac{1}{E} \rightarrow
\Omega_+(E) =  e^{2 i \theta(E)} + O(\frac{1}{E})  
\label{r21}
\eeq
\no
which implies 
\beq
\CF(E)  =    1 + a_0 b_0  e^{2 i \theta(E)} 
+ \left( \frac{a_0 b_0 }{2} \right)^2 +  O(\frac{1}{E})
\label{r22} 
\eeq
\no
This Jost function has zeros on the real axis, up to order $1/E$,
provided 
\beq
\ep = \frac{a_0 b_0}{2} = \pm 1 \Longrightarrow
\CF(E)  =  2(  1 + \ep \;   e^{2 i \theta(E)})  
+  O(\frac{1}{E})
\label{r23} 
\eeq
\no
The choice $\ep = - 1$ reproduces the smooth part of the
Riemann formula (\ref{s6}) since,
\beq
\ep = - 1 \Longrightarrow  1
- e^{2 i \theta(E)} = 
1 - e^{2 \pi i \langle \CN(E) \rangle } =  0
\label{r24} 
\eeq
\no
where $E$ is the average position of the zeros. 
On the other hand the choice $\ep = 1$ leads to
\beq
\ep = 1 \Longrightarrow  1
+ e^{2 i \theta(E)} = 0 \Longrightarrow
\cos \theta(E) = 0 
\label{r25} 
\eeq
\no
so that the number of zeros in the interval
$(0,E)$ is given by
\beq
\CN_\sm(E) = \frac{\theta(E)}{\pi} + \frac{3}{2}
\label{r26} 
\eeq
\no 
which gives a better numerical approximation
than the term $\langle \CN(E) \rangle$ that appears in the 
exact Riemann formula (\ref{s6}) 
(see also fig.2). In the case of the sine boundary
function (\ref{r2}) one similarly  obtains
the smooth part of the zeros of the odd
Dirichlet L-functions. 

In summary, we have shown that the 
semiclassical BK boundary conditions
have a quantum counterpart 
in terms of the  boundary wave functions $\psi_{a,b}$,
and that the average  Riemann
zeros become asymptotically  bound states
of the model or more appropiately resonances.

\section{The quantum model of the Riemann zeros}

In section II we showed how to incorporate
the fluctuations of the energy levels in the
heuristic $xp$ model by means of the
functions $p_\cl(x)$ and $x_\cl(p)$ which
define the boundaries of the allowed
phase space. These functions are given by eq.(\ref{s23}) 
in terms of the density of the fluctuation part of the energy levels. 
In the quantum model the functions
 $p_\cl(x)$ and $x_\cl(p)$ are represented
by the wave functions $\psi_a$ and $\psi_b$.
Hence it is natural to impose the following 
conditions 
\barray 
\left( \log \frac{|\hp|}{l_p} + \pi \;  n'_{\rm fl}(H_0) 
\right) |\psi_a \rangle = & 0 &  
\label{q1} \\
\left( \log \frac{\hx}{l_x} + \pi \;  n'_{\rm fl}(H_0)
\right) |\psi_b \rangle = & 0 &
\label{q2}
\earray 
\no
where $n'_{\rm fl}(E) = d n_{\rm fl}(E)/dE$
and $H_0$ is the no interacting Hamiltonian
(\ref{t1}). The hat over $x$ and $p$ stress
the fact that they are operators. 
Eqs.(\ref{q1}) and  (\ref{q2})  can be taken as the definition
of the boundary wave functions. 
To solve these eqs. let us  write them as
\barray 
\left( \log |\hp| + \lambda_p +  \pi \;  n'_{\rm fl}(H_0) 
\right) |\psi_a \rangle =& 0 &,   
\label{q3} \\
\left( \log \hx + \lambda_x + \pi \;  n'_{\rm fl}(H_0)
\right) |\psi_b \rangle = & 0 &, 
\label{q4} \\
  \lambda_p = - \log l_p, \;  \lambda_x = - \log l_x & & 
\label{q5} 
\earray 
\no
It is convenient to expand the states $|\psi_{a,b} \rangle$
in the basis (\ref{t3}) 
\beq
|\psi_{a,b} \rangle = \int_{- \infty}^\infty dE  \; 
\psi_{a,b}(E) \; |\phi_E \rangle, \quad
\langle x | \phi_E \rangle = \frac{1}{\sqrt{ 2 \pi}}
\frac{1}{x^{1/2 - i E}} 
\label{q6} 
\eeq
\no 
Let us first consider eq.(\ref{q4}) which in the basis
(\ref{q6}) becomes
\beq
\int_{- \infty}^\infty  dE' \; 
\langle \phi_E| \log \hx | \phi_{E'} \rangle \;
\psi_b(E') + ( \lambda_x + \pi \; n'_{\rm fl}(E) ) \psi_b(E) = 0 
\label{q7} 
\eeq
\no
The matrix elements of the operator $\log \hx$ 
can be readily computed, 
\beq
\langle \phi_E| \log \hx | \phi_{E'} \rangle \; = 
- i \; \delta'(E' - E) 
\label{q8} 
\eeq
\no
which replaced in (\ref{q7}) and upon integration yields 
\beq
i \frac{d \psi_b(E)}{dE}  + ( \lambda_x + \pi \;
  n'_{\rm fl}(E) ) \psi_b(E) = 0 
\label{q9} 
\eeq
\no
 The solution of (\ref{q9}) is simply
\beq
\psi_b(E) = \psi_{b,0}  \; e^{ i 
( \lambda_x \; E + \pi \;  n_{\rm fl}(E) ) } 
\label{q10} 
\eeq
\no
where $\psi_{b,0}$ is an integration constant. 
The $x$-space representation of $\psi_b$ 
follows from (\ref{q10}) and (\ref{q6})
\beq
\psi_b(x)  =   \int_{- \infty}^\infty dE \;
\psi_b(E) \;  \phi_E(x) 
=   \psi_{b,0}  
\int_{- \infty}^\infty \frac{dE}{\sqrt{2 \pi}} \;
 e^{ i ( \lambda_x \; E +  \pi \; n_{\rm fl}(E) ) } 
x^{-1/2 + i E} 
\label{q11}
\eeq
\no
Recalling that  $ \psi_b(x) = b(x)/\sqrt{x}$ one gets
\beq
b(x) =  \psi_{b,0}  
\int_{- \infty}^\infty \frac{dE}{\sqrt{2 \pi}} \;
 e^{ i ( \lambda_x \; E +  \pi \;  n_{\rm fl}(E) ) } 
x^{ i E}
\label{q12} 
\eeq
\no
Observing that $b(x)$ is related
to its Fourier transform $\hb(E)$, as
\beq
b(x) =  
\int_{- \infty}^\infty \frac{dE}{2 \pi} \;
\hb(E) \;  x^{- i E}
\label{q13} 
\eeq
\no
one finally obtains 
\beq
\hb(E) = \sqrt{2 \pi} \; \psi_{b,0} \; 
 e^{ - i ( \lambda_x \; E +   \pi \; n_{\rm fl}(E))}  
\label{q14} 
\eeq
\no
where we assumed that $n_{\rm fl}(E)$ is an odd
function of $E$. If  $n_{\rm fl}(E)=0$, 
eq.(\ref{q14}) reproduces (\ref{r10}), i.e.  
\beq
n_{\rm fl}(E) = 0 \Longrightarrow
\hb(E) = \sqrt{2 \pi} \; \psi_{b,0} \; l_x^{i E} = 
b_0 \; l_x^{ i E}
\label{q15} 
\eeq
\no
To simplify the notations we shall  write (\ref{q14})
as
\beq
\hb(E) = b_0 \;  l_x^{ i E} \; 
 e^{ - i   \pi   n_{\rm fl}(E)}  
\label{q16} 
\eeq
\no
Let us now solve the condition (\ref{q3})
for the wave function $\psi_a$. 
We first need to define the operator $\log |\hp|$
acting in the Hilbert space 
expanded by the functions $\phi_E \; (E \in \Rmath)$. 
In this respect it  is worth to remember 
that the operator $\hp = - i d/dx$ is self-adjoint 
in the real line $(- \infty, \infty)$ and in the 
finite intervals $(a,b)$, but not
in the half-line $(0, \infty)$ \cite{self}. However, the operator
$\hp^2$ admits infinitely many self-adjoint extensions
in the half-line provide the wave functions
satisfy the boundary condition 
\beq
\psi'(0) =  \kappa \; \psi(0) 
\label{q17} 
\eeq
\no
where $\kappa \in \Rmath \cup  \infty$. 
We shall confine ourselves to the cases 
where $\kappa = 0$  and $\infty$, which correspond
to the von Neumann and Dirichlet BC's respectively, 
\barray 
\kappa & = &  0 \rightarrow \psi' (0) = 0, 
\label{q18} \\
\kappa &  = &  \infty  \rightarrow \psi (0) = 0
\nonumber 
\earray 
\no
The corresponding eigenstates of the operator $\hp^2$ 
with eigenvalues $p^2$ read
\beq
\left\{ 
\begin{array}{l}
\chi_p^+ \\
\chi_p^-  \\
\end{array}
\right.
= \sqrt{ \frac{2}{\pi}} \times
\left\{ 
\begin{array}{ll}
\cos( p x)  &  (p>0) \\
\sin( p x) & (p>0)  \\
\end{array}
\right.
\label{q19}
\eeq
\no
These
basis are complete in the space of functions defined in 
 $(x >0)$, i.e.
\beq
\int_0^\infty dp \;  (\chi_p^\eta(x))^* \; \chi_p^\eta(x')
= \delta(x - x'),\;\; x,x' >0, \;\; \eta = \pm 
\label{q20} 
\eeq
\no
The operator $\log | \hp |$ will be defined as 
$\frac{1}{2} \log \hp^2$, and therefore
admits the same self-adjoint extensions
as  $\hp^2$. The analogue of eq.(\ref{q7}) reads now
\beq
\int_{- \infty}^\infty  dE' \; 
\langle \phi_E| \log |\hp| \;  | \phi_{E'} \rangle \;
\psi_a(E') + ( \lambda_p +  \pi \;  n'_{\rm fl}(E) ) \psi_a(E) = 0 
\label{q21} 
\eeq
\no
The matrix elements of $\log |\hp|$ 
can be computed introducing the resolution
of the identity in  the basis
(\ref{q19}), 
\beq
\langle \phi_E| \log |\hp| \;  | \phi_{E'} \rangle 
= \int_{0}^\infty dp \; \log p \; 
\langle \phi_E| \chi_p^\eta \rangle \langle  \chi_p^\eta | \phi_{E'} \rangle 
\label{q22} 
\eeq
\no
where the overlap of the eigenstates of $\hp^2$
and $H_0$ are 
\beq
\langle  \chi_p^\pm  | \phi_{E} \rangle
= 
\int_0^\infty \frac{dx}{\pi} \; x^{-\frac{1}{2} + i E} 
\times 
\left\{
\begin{array}{c}
\cos( p x) \\
\sin(p x) \\
\end{array}
\right. 
\label{q23} 
\eeq
\no
These integrals were already computed
in eq.(\ref{r6}), and the result is 
\beq
\langle  \chi_p^\pm  | \phi_{E} \rangle
= \frac{(2 \pi)^{-1/2 + i E}}{p^{1/2 + i E}} \; e^{2 i \theta_\pm(E)}
\label{q24} 
\eeq
\no
Plugging this eq. into (\ref{q22}), and performing the
integral gives 
\beq
\langle \phi_E| \log |\hp| \;  | \phi_{E'} \rangle 
= i \; \delta'(E' - E) ( 2 \pi)^{i (E' - E)}
\; e^{2 i (\theta_\eta(E') - \theta_\eta(E))}  
\label{q25} 
\eeq
\no
which introduced in (\ref{q21}) yields
a differential equation whose solution is
\beq
\psi_{a_\eta}(E) = \psi_{a,0}  \;(2 \pi)^{- i E}  e^{- i 
( \lambda_p \; E + \pi  n_{\rm fl}(E) + 2 \theta_\eta(E) ) } 
\label{q26} 
\eeq
\no
The function $\psi_a(x)$ reads 
\beq
 \psi_{a_\eta}(x)  =   \int_{- \infty}^\infty dE \;
\psi_{a_\eta}(E) \;  \phi_E(x) =   \psi_{a,0}  
\int_{- \infty}^\infty \frac{dE}{\sqrt{2 \pi}} 
(2 \pi)^{- i E} 
 e^{ - i ( \lambda_p \; E +  \pi  n_{\rm fl}(E) +  2 \theta_\eta(E) ) } 
x^{-1/2 + i E} 
\label{q27}
\eeq
\no
while
\beq
a_\eta(x)  =   \psi_{a,0}  
\int_{- \infty}^\infty \frac{dE}{\sqrt{2 \pi}} 
(2 \pi)^{- i E} 
 e^{ - i ( \lambda_p \; E +  \pi  n_{\rm fl}(E) +  2 \theta_\eta(E) ) } 
x^{i E} 
\label{q28}
\eeq
\no
whose Fourier transform is
\beq
\ha_\eta(E)  =   \psi_{a,0} 
(2 \pi)^{ 1/2 + i E} 
 e^{  i ( \lambda_p \; E +  \pi  n_{\rm fl}(E) +  2 \theta_\eta(E) ) } 
\label{q29}
\eeq
\no
If there are no fluctuations, eq.(\ref{q29})
reduces to
\beq
n_{\rm fl}(E) = 0 \Longrightarrow
 \ha_\eta(E)  =  \sqrt{2 \pi} \psi_{a,0} 
\left( \frac{2 \pi}{l_p} \right)^{i E} 
e^{ 2 i \theta_\eta(E) ) } 
\label{q30}
\eeq
\no
which coincides with eq.(\ref{r10}). To 
simplify notations we shall write (\ref{q29})
as
\beq
\ha_\eta(E)  =   a_0  
\left( \frac{2 \pi}{l_p} \right)^{ i E}
 e^{  i (  \pi  n_{\rm fl}(E) +  2 \theta_\eta(E) ) } 
\label{q31}
\eeq
\no
The two solutions (\ref{q16}) and  (\ref{q31})
satisfy the duality relation (\ref{r9})
and hence the wave functions $\psi_{a_\pm}(x)$
is the cosine or sine Fourier transform
of $\psi_b(x)$ ( see eq. (\ref{r3})).

Having found the boundary wave functions 
for generic fluctuations we turn into the 
computation of the corresponding Jost
function.
The basic products of the $\ha$ and $\hb$
functions needed to find the $S_{f,g}$
functions are similar to eqs.(\ref{r11}), 
\barray
& \ha_\pm (E) \;  \ha_\pm (-E) = a_0^2, & \nonumber \\
&  \hb(E) \;  \hb(-E) = b_0^2 & 
\label{q32} \\
& \ha_\pm (E) \;  \hb(-E) = a_0 b_0 \; e^{2 i ( \theta_\pm(E) + 
\pi n_{\rm fl}(E))  } 
& \nonumber \\
& \hb (E) \;  \ha_\pm(-E) = a_0 b_0 \; e^{- 2 i (\theta_\pm(E) + 
\pi n_{\rm fl}(E)} 
& \nonumber
\earray 
\no
The diagonal terms of $S_{f,g}$ 
are the same as in eq.(\ref{r12}), i.e.
\beq
S_{a_\pm, a_\pm}(E) = \frac{a_0^2}{2},
\qquad 
S_{b,b}(E) = \frac{b_0^2}{2} 
\label{q33}
\eeq
\no
while the evaluation of 
the off-diagonal terms depends on the analytic properties
of the function $e^{ 2 \pi i \; n_\pm(E)}$ where
\beq
n_\pm(E) \equiv 
\frac{\theta_\pm(E)}{\pi}  + n_{\rm fl}(E)) 
\label{q34}
\eeq
\no
This definition is strongly reminiscent
of the Riemann formula (\ref{s6}), with
$n_\pm(E)$ playing the role of $\CN_R(E)$, 
and $n_{\rm fl}(E)$ that of $\CN_{\rm fl}(E)$.
However,  we must keep in mind that 
 $\CN_R(E)$ is a step function
while we expect $n_\pm(E)$ to be a continuous
interpolating function between the zeros. 
The value of $S_{a_\pm, b}$ is given
by the integral 
\beq
S_{a_\pm,b}(E)  =  \frac{a_0 b_0}{2} 
\left[ e^{2 \pi  i n_\pm(E)} + 
P \int_{- \infty}^\infty \frac{dt}{\pi i} \frac{ e^{2 \pi i 
n_\pm(t)} }{t - E} 
\right]
\label{q35}
\eeq
\no
We shall make the asumption that $e^{2 \pi  i n_\pm(E)}$
is an analytic function in the upper half plane
which goes to zero as $|E| \rightarrow \infty$. 
In this case the Cauchy integral on the RHS of (\ref{q35})
is equal to 
$e^{2 \pi  i n_\pm(E)}$
and one finds
\beq
S_{a_\pm,b}(E)  =  a_0 b_0 \;  e^{2 \pi  i n_\pm(E)} 
\label{q36}
\eeq
\no
Similarly $S_{b, a_\pm}$ vanishes so that the
Jost function reduces to 
\beq
\CF(E)  =    1 + a_0 b_0 \;   e^{2 \pi i  n_\pm(E)} 
+ \left( \frac{a_0 b_0 }{2} \right)^2 
\label{q37} 
\eeq
\no
and  under the usual choice 
\beq
\ep = \frac{a_0 b_0}{2} = \pm 1 \Longrightarrow
\CF(E)  =  2(  1 + \ep \;   e^{2 \pi  i n_\pm(E)})  
\label{q38} 
\eeq
\no
When $n_{\rm fl}=0$ the results of the previous subsection
showed that $\ep = 1$ gives a better numerical 
estimate to the smooth part of the zeros. In the sequel
we shall also make that choice which implies that
the number  of zeros of $\CF(E)$ in the interval $(0,E)$ 
is 
\beq
\CN_\QM(E) =\CN_\sm(E) + n_\fl(E) = n_\pm(E) + \frac{3}{2} 
\label{q39} 
\eeq
\no
where $\CN_\sm(E)$ was defined in (\ref{r26}) for the
particular case of the zeta function $\zeta(s)$,
which corresponds to $n_+(E)$. Equation 
(\ref{q39}) agrees asymptotically with the semiclassical
formula (\ref{s20-1}), which  confirms  the ansatz made 
for  the states $\psi_a$ and $\psi_b$. 

\subsection*{The connection with the Riemann-Siegel formula}

The next problem is  to find 
the function $n_\fl(E)$, and therefore $\CN_\QM(E)$,
which gives the exact location of the Riemann zeros. 
Let us consider the case of the zeta function 
with the following choices of parameters
\beq
\eta = +, \;\; \ep = 1, \;\;  a_0 = b_0 =
\sqrt{2}, \;\; l_x = 1, \;\; l_p = 2 \pi  
\label{q40} 
\eeq
\no  
which correspond to the potentials (recall (\ref{q31}) 
and (\ref{q16})) 
\barray 
\ha(t) &  = &  e^{i ( 2 \theta(t) + \pi n_\fl(t) )} 
=  e^{i ( \theta(t) + \pi n(t) )} 
\label{q41} \\
\hb(t) &  = &  e^{-i \pi n_\fl(t) )} 
=  e^{i (\theta(t) - \pi n(t) )} 
\nonumber 
\earray 
\no
where we skip a common factor $\sqrt{2}$ 
and denote $n(E) \equiv n_+(E)$. These two
functions are interchanged under the  
transformation 
\barray 
\ha(t) &  \rightarrow  &
  e^{2 i \theta(t)} \;  \ha(-t) = \hb(t) 
\label{q42} \\
\hb(t) &  \rightarrow  &
  e^{2 i  \theta(t)} \;  \hb(-t) = \ha(t) 
\nonumber 
\earray 
\no
so that their sum is left invariant,  
\beq
\ha(t) + \hb(t) \rightarrow 
e^{2 i \theta(t)} (\ha(-t) + \hb(-t))
=  \ha(t) + \hb(t)
\label{q43}
\eeq
\no
The functional relation satisfied by the zeta
function implies 
\beq
\zeta(1/2 - i t)  \rightarrow 
e^{2 i \theta(t)} \zeta(1/2 + i t) = \zeta(1/2 - i t)
\label{q44}
\eeq
\no 
which suggests to relate $\ha + \hb$ and $\zeta$ as
\beq
\zeta(1/2 - i t)  = \rho(t) \; 
 (\ha(t) + \hb(t))
\label{q45}
\eeq
\no
where $\rho(t)$ is a proportionally factor. 
Using eqs.(\ref{q42}) into (\ref{q45})
yields
\beq
\zeta(1/2 - i t)  = 2 \; \rho(t) \; e^{i \theta(t)} 
\cos(\pi n(t)) 
\label{q46}
\eeq
\no
This formula can be compared with the 
parametrization  of the zeta function
in terms of the Riemann-Siegel zeta function
$Z(t)$ and its phase $\theta(t)$, 
\beq
\zeta(1/2 - i t)  = Z(t) \;  e^{i \theta(t)}
\label{q47}
\eeq
\no
which leads to, 
\beq
Z(t) = 2 \; \rho(t) \;  \cos(\pi n(t)) 
\label{q48}
\eeq
\no
This equation is rather interesting since
it implies that the zeros of $\cos(\pi n(t))$,
which give the bound states of the QM model,
are also zeros of $Z(t)$, of course if 
$\rho(t)$ does not have poles at those values. 
Viceversa, the zeros of $Z(t)$ can be zeros
either of  $\cos(\pi n(t))$, or of $\rho(t)$, or both. 
The latter possibility would be absent if the Rieman
zeros are simple, as it is expected to be the case.

\begin{figure}[t!]
\begin{center}
\includegraphics[height= 7.0 cm,angle= 0]{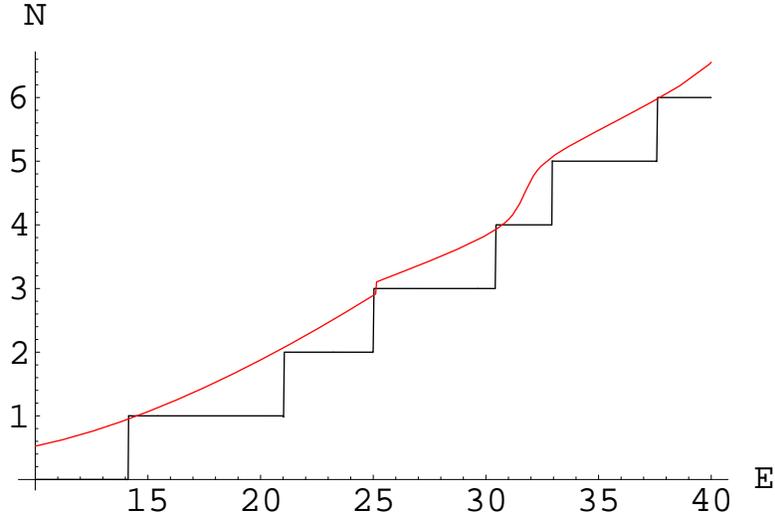}
\end{center}
\caption{In black: $\CN_R(E)$, in red: 
$\CN_\QM(E)$ in the interval $(10,40)$ 
}
\label{nqm}
\end{figure}

A first hint on the structure of the functions
$\rho(t)$ and  $\cos(\pi n(t))$ can be obtained
using the Riemann-Siegel formula for $Z(t)$,
\beq
Z(t) = 2 \sum_{n=1}^{\nu(t)} n^{-1/2}
\cos( \theta(t) - t \log n) + R(t), \;\; \nu(t) = 
\left[\sqrt{ \frac{t}{2 \pi} } \right] 
\label{q49}
\eeq
\no
where  $[x]$ the integer part of $x$ and 
$R(t)$ is a reminder of order $t^{-1/4}$. 
Combining the last two equations one finds
\barray 
Z(t) & = & 
2 \rho(t) \; \left[  \cos \theta(t) \; \cos( \pi n_\fl(t) )
-  \sin \theta(t) \; \sin( \pi n_\fl(t) ) \right] 
\label{q50}
\\
& \sim & 2 \left[  \cos \theta(t)
\sum_{n=1}^{\nu(t)} \frac{ \cos( t \log n)}{n^{1/2}} 
+  \sin \theta(t)
\sum_{n=1}^{\nu(t)} \frac{ \sin( t \log n)}{n^{1/2}} 
 \right]
\nonumber 
\earray 
\no
which suggests  the following identifications
\barray 
\rho(t)  \cos( \pi n_\fl(t) )  & \sim &  
\sum_{n=1}^{\nu(t)} \frac{ \cos( t \log n)}{n^{1/2}}
\label{q51}
\\
\rho(t)  \sin( \pi n_\fl(t) )  & \sim  &  
- \sum_{n=1}^{\nu(t)} \frac{ \sin( t \log n)}{n^{1/2}}
\nonumber 
\earray 
\no
that can be  combined into
\beq
f(t) \equiv  \rho(t)  e^{i \pi n_\fl(t)} \sim 
\sum_{n=1}^{\nu(t)} \frac{1}{n^{1/2+ i t}}
\label{q52}
\eeq 
\no
The fluctuation function 
$n_\fl(t)$ is then given by the phase of $f(t)$, i.e. 
\beq
n_\fl(t) =  \frac{1}{\pi} \Im \log  f(t) 
\label{q53}
\eeq 
\no
In fig. \ref{nqm} we plot the values of 
$\CN_{QM}(t)$ that correspond to the approximate formula
(\ref{q52}), which shows an excelent agreement
with the Riemann formula (\ref{s6}). This is expected
from the fact that the main term of the 
Riemann-Siegel formula already gives accurate
results for the lowest Riemann zeros. For higher
zeros one has to compute more
terms of the reminder $R(t)$ depending on the
desired accuracy. 
Observe that $\CN_{QM}(t)$ is a smooth function,
except for some jumps at higher values of $t$ (not shown in fig. \ref{nqm}) 
due to the approximation made, 
unlike $\CN_R(t)$, which is a step function.

\begin{figure}[t!]
\begin{center}
\includegraphics[height= 7.0 cm,angle= 0]{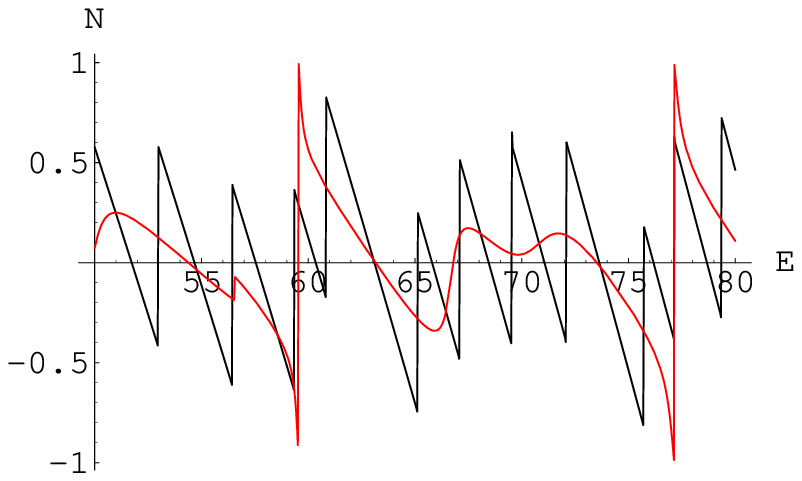}
\end{center}
\caption{In black: $\CN_{\rm fl}(E)$, in red: 
$n_\fl(E)$ in the interval $(50,80)$ 
}
\label{nfl}
\end{figure}

\no 
In fig. \ref{nfl} we plot the  values of (\ref{q53}) 
together with those
of the fluctuation part of the Riemann
formula (\ref{s6}), i.e.
\beq
\CN_{\rm fl}(t) =  \frac{1}{\pi} \Im \log \zeta \left( 
\frac{1}{2} + i t \right)
\label{q54}
\eeq

\no
The jumps in  $\CN_{\rm fl}(t)$ correspond to the Riemann
zeros, while those
of $n_\fl(t)$ correspond, either to  jumps of the function
$\nu(t)$ appearing in the Riemann Siegel formula
(\ref{q49}), or to those points where the curve 
$f(t)$ cuts the negative real axis in the complex plane. 

\begin{figure}[t!]
\begin{center}
\hbox{\includegraphics[height= 5.0 cm]{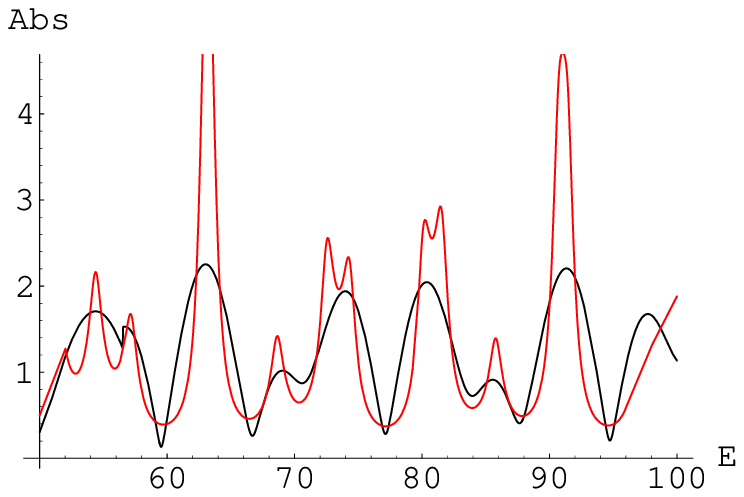}
\includegraphics[height = 5.0 cm]{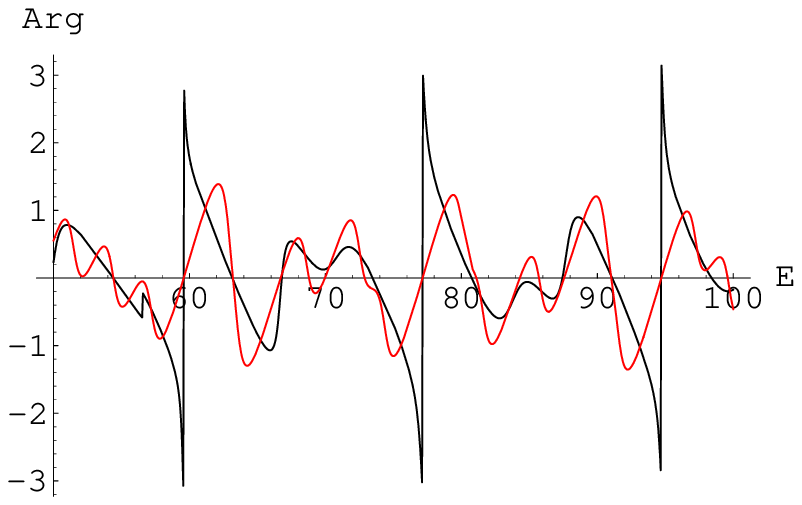}}
\end{center}
\caption{Left: in  black: $|f(E)|$, in red: 
$|\zeta_\E(1/2 + i E)|$ in the interval $(50,100)$.
Right:  in  black: ${\rm Arg} \;  f(E)$, in red: 
${\rm Arg} \; \zeta_\E(1/2 + i E)$. 
}
\label{eulerRS}
\end{figure}

We gave in section II a formal expression of eq.(\ref{q54}) 
in terms of prime numbers, eq. 
(\ref{s8}), which resembles the fluctuation part (\ref{s9})
of a quantum chaotic system. Eq.(\ref{s8}) is based
on the Euler product formula (\ref{s7-2}) 
which is not valid in the case where
$s = 1/2 + it$, since $\Re \; s > 1$ 
for convergence of the infinite product. The Euler product
formula does not apply to the truncated sum (\ref{q52}),
however we shall naively try to establish a relationship. 
Let us denote by $p_n$ the $n^{\rm th}$-prime number, e.g. 
$p_1=2, p_2 = 3$, etc, and by $\Pi(x)$ the number of primes
less or equal to $x$. The sum (\ref{q52}) involves all 
integers up to $\nu(t)$, which can be expressed as
products of the first $\mu(t)$ prime numbers where
\beq
\mu(t) =  \Pi( \nu(t)), \;\; p_{\mu(t)} = {\rm inf} \;\{ p
\} < \nu(t)  
\label{q55}
\eeq
\no
Using these functions we define a truncated
Euler product as
\beq
\zeta_\E(1/2 + i t)  \equiv \prod_{n = 1}^{\mu(t)} 
\frac{1}{1 - p_n^{-1/2 - i t}} 
\label{q56}
\eeq
\no
It is easy to see that $\zeta_\E(1/2 + i t)$
is not equal to  $f(t)$, for there
are terms in (\ref{q56})
which do not appear in (\ref{q52}), although 
all the terms appearing in the latter sum also appear
in the former product. The point is that a numerical
comparison of these 
two functions shows a qualitative agreement as depicted
in fig. \ref{eulerRS}. Indeed, the minima and maxima of their
absolute value are located around the same points, and the same
happens for the zeros of their arguments. The conclusion we
draw from these heuristic considerations is that the function $f(t)$
contains some sort of information related to the primes numbers
although not in the form of an Euler product formula as is the case
of $\zeta_\E(1/2 + i t)$.  It would be interesting to investigate
the consequences of this results from the point of view
of Quantum Chaos.

\subsection*{The Berry-Keating formula of $Z(t)$}

The main term of the 
Riemann-Siegel formula (\ref{q49}) is not analytic 
in $t$ due to the discontinuity in the main sum.
This problem was solved by Berry and Keating who
found an alternative expression for $Z(t)$ \cite{BK3}. 
The formula is 
\beq
Z(t) = \sum_{n=1}^\infty ( T_n(t) + T_n(-t))
\label{bk1}
\eeq
\no
where 
\barray 
T_n(t)&  = & T_n^*(-t)=
  \frac{e^{i \; \theta(t)}}{n^{1/2 + i t}} \;  \beta_n(t) 
\label{bk2} 
\\
  \beta_n(t)& = & \frac{1}{2 \pi i} \int_{C_-} 
\frac{dz}{z} e^{- z^2 \; K^2/(2 |t|)} \;
e^{ i [ \theta(z+t) - \theta(t) - z \log n]} 
\nonumber 
\earray 
\no
and $C_-$ is an integration contour in the lower
half plane with $\Im \; < -1/2$ that avoids
a cut starting at the brach point $z = - t - i/2$. 
The constant $K$ in (\ref{bk2}) can be choosen at will
and it is related to the number of terms of the RS
formula that has been smoothed for large values of $t$. 
Using eq.(\ref{bk1}) one can write the zeta function as
\beq
\zeta(1/2 - i t) = e^{2 i \theta(t)}
 \sum_{n=1}^\infty \frac{ \beta_n(t)}{n^{1/2 + i t}} 
+  \sum_{n=1}^\infty \frac{ \beta_n(-t)}{n^{1/2 - i t}} 
\label{bk3}
\eeq
\no
which can be compared with (\ref{q45}) obtaining 
\beq
f(t) = \rho(t) e^{ i \pi n_\fl(t)} = 
 \sum_{n=1}^\infty \frac{ \beta_n(t)}{n^{1/2 + i t}} 
\label{bk4}
\eeq
\no
so that (\ref{bk3}) can be written as 
\beq
\zeta(1/2 - i t) = e^{2 i \theta(t)} \; f(t) + f(-t) 
\label{bk3b}
\eeq
\no
Eq.(\ref{bk4})  gives an exact expression of $f(t)$,
which is  in fact a smooth version of (\ref{q52}).
Berry and Keating also found a series for $Z(t)$
which improves the RS series. The first term of that
series corresponds to the following value of the 
$\beta_n(t)$ functions  
\barray 
\beta_n^{(0)}(t)&  = & \frac{1}{2} {\rm Erfc}
\left( \frac{\xi(n,t)}{ Q(K,t)} \sqrt{t/2} \right) 
\label{bk5} 
\\
\xi(n,t) & = & \log n - \theta'(t), 
\;\; 
Q^2(K,t) = K^2 - i t \theta''(t) 
\nonumber 
\earray 
\no
where $Erfc$ is the complementary error function. 
Using these formulas one can find a better
numerical evaluation of the functions
$\CN_{QM}(t)$ and $n_\fl(t)$.

It is perhaps worth to mention that eq.(\ref{bk3b}), 
with the approximate value of $f(t)$ given
by (\ref{q52}), is a particular case of the so called
aproximate functional relation due to Hardy and Littlewood 
\cite{Edwards,Titchmarsh2}
\beq
\zeta(s) = \sum_{n \leq x} n^{-s} + \pi^{s- 1/2} \; 
\frac{ \Gamma((1-s)/2)}{ \Gamma(s/2)} 
\sum_{n \leq y} n^{1-s} + O(x^{ - \sigma}) + O( |t|^{1/2 - \sigma} 
y^{\sigma -1}) 
\label{bk3c}
\eeq
where $s = \sigma + i t$, $|t| = 2 \pi x y$, $0 < \sigma < 1$.  
Recalling that in our model $t$ is the energy $E$, then 
equation $|t| = 2 \pi x y$ becomes the hyperbola $|E| = x p$ with
$p = 2 \pi y = l_p y$ so that the sums in (\ref{bk3c}) 
run over the integer values of the positions and momenta
in units of $l_x$ and $l_p$ respectively. Eq.(\ref{bk3c})
also suggests that the case where $\sigma \neq 1/2$
could be related to the non hermitean Hamiltonian
$H_0 = (xp + p x)/2 - i (\sigma-1/2)$ whose right (resp. left) 
eigenfunctions are given by $1/x^{\sigma - i E}$ ( resp. 
 $1/x^{1- \sigma - i E}$).

On more general grounds, we would  like to mention two important points.
First is that one still needs to show that the function 
 $n(t)$, defined
in eq. (\ref{q34}), is such that $e^{2 \pi i n(t)}$
is analytic in the upper-half plane and that 
it goes to zero as $|t| \rightarrow \infty$, so that
the Jost function is indeed given by eq.(\ref{q39}),
as we have assumed so far.  Second, and related to the latter point,  is that 
that the function $n_\fl(t)$ is well defined provided $f(t)$ does not vanish
for $t$ real, in which case (\ref{bk3b}) reads also 
\beq
\zeta(1/2 - i t) = f(-t) \left( 1 + e^{2 i \theta(t)} \frac{ f(t)}{f(-t)} \right) 
= f(-t) \; \F(t) 
\label{bk6}
\eeq
\no 
which shows that  our construction of a QM model of the Riemann zeros
 relies on the absence
of zeros of the function $f(t)$ on the critical line. These zeros were investigated
by Bombieri long ago in an attempt to improve the existing lower bounds for the number
of Riemann zeros on the critical line  \cite{Bombieri2}. In this regard
our results  give further support, but not a proof, to the RH. 
As suggested in \cite{Sierra2,Sierra3} that proof would follow
if the  zeta function $\zeta(1/2 - i t)$ can be realized 
as the Jost function of a QM model of the sort discussed so far,  due
to its special analyticity properties. Eq.(\ref{bk6}) gives
a partial realization of this idea but the function $f(t)$ lacks 
of a  physical interpretation so far. 
The latter approach is analogue to the ones proposed in the past 
by several authors where the zeta function
gives the scattering phase shift of some quantum
mechanical model, particularly on the line $\Re \; s = 1$
\cite{Faddeev,Lax,G2,Joffily,BKL}. 

Another important question is: where are the prime numbers in 
our construction? As suggested by the Quantum Chaos scenario,
the prime numbers may well be classical objects hidden
in the quantum model, so the next question is: what is the classical
limit of the Hamiltonian?. The free part is of course given 
by $xp$, but the interacting part is an antisymmetric matrix
with no obvious classical version. The existence of such a
classical Hamiltonian may help to answer the {\em prime}
question but it may also lead to a real physical
realization of the model. 
Work along this direction is under progress \cite{SP}.

\section*{Acknowledgments}
I wish to thank for discussions  M. Asorey, M. Berry, 
L.J. Boya,  J. Garc\'{\i}a-Esteve, J. Keating,  J.I. Latorre, A. LeClair, J. Links, 
M.A. Mart\'{\i}n-Delgado, G. Mussardo, 
J. Rodr\'{\i}guez-Laguna and  P.K. Townsend.  
This work was supported by the 
CICYT of Spain
under the contracts FIS2004-04885.  I also acknowledge 
ESF Science Programme INSTANS 2005-2010.

\section{Appendix A: Wave functions and norms}

\def\om{{\omega}}

In this appendix we shall derive alternative expressions
of the eigenfunctions of the model and compute their norm.
Let us start from eq.(\ref{u7}) for the eigenfunctions
of the Hamiltonian (\ref{t5}), 
\beq
\psi_E(q)   =
    e^{- (1/2 - i E) q }
\left[ 
C_0 + \int_{-\infty}^q  dq' \;  e^{- i E q'}
( B \; a(q') - A \;  b(q'))  
\right] 
\label{A1}
\eeq
Replacing $a(q)$ and $b(q)$ by their Fourier transform, and using eq.(\ref{a3})
 one finds
\beq
 \int_{-\infty}^q  dq' \;  e^{- i E q'} a(q') = 
 \frac{\ha(-E)}{2} +  e^{- i q E} \int_{-\infty}^\infty 
\frac{d \om}{2 \pi i}  \; 
 \frac{  e^{i q \om}  \; \ha(-\om)}{ \om - E} 
\label{A2}
\eeq
and a similar expression for the integral of $b(q)$. All the 
 singular integrals appearing in this appendix 
must be understood in the Cauchy sense. 
 Plugging the latter
expressions into (\ref{A1}) yields
\beq
\psi_E(q)   =
    e^{- (1/2 - i E) q }
\left[ 
C_0 +   \frac{1}{2} (B \ha(-E) - A \hb(-E) ) + e^{- i q E} 
\int_{-\infty}^{\infty} \frac{d \om}{2 \pi i} \;  e^{i q \om} \; 
\frac{ B \ha(-\om) - A \hb( - \om)}{ \om - E}
\right] 
\label{A3}
\eeq
Using eqs.(\ref{u9}),  (\ref{u28})    and (\ref{u29}), 
the first term in the RHS becomes
\beq
C_0 +   \frac{1}{2} (B \ha(-E) - A \hb(-E) )  = 
\frac{ C_0 + C_\infty}{2} = \Re \; \F(E) 
\label{A4}
\eeq
so that $\psi(x)$ is given by 
\beq
\psi_E(x)   = \frac{ \Re \F(E)}{x^{1/2 - i E}} + 
 \int_{-\infty}^{\infty} \frac{d \om}{2 \pi i} \; x^{-1/2 + i \om} \; 
\frac{ B(E) \ha(-\om) - A(E)  \hb( - \om)}{ \om - E}
\label{A5}
\eeq
where $A(E)$ and $B(E)$ are given by the eqs.(\ref{u28}) and
(\ref{u29}). The function (\ref{A5}) can also be expanded
in the basis (\ref{t3}) of eigenfunctions of $H_0$, i.e. 
\beq
|\psi_E \rangle = \int_{- \infty}^\infty 
d\om  \; \psi_E(\om) \; |\phi_{\om} \rangle 
\label{A6}
\eeq
namely
\beq
\psi_E (x)  = \int_{- \infty}^\infty 
d\om  \; \psi_E(\om) \; \frac{ x^{-1/2 + i \om}}{\sqrt{2 \pi}}
\label{A7}
\eeq
\no
The result is
\beq
\psi_E(\om) = \sqrt{2 \pi} \;
  \delta(E-\om) \; \Re \; \CF(E)  + \frac{1}{\sqrt{2 \pi} i} 
\frac{ B(E) \ha(-\om) - A(E)  \hb( - \om)}{ \om - E}
\label{A8}
\eeq
\no
which shows that the delocalized states, i.e.
$\CF(E) \neq 0$,  have to be normalized in the
distributional sense, while the
localized states, i.e. $\CF(E_m) = 0$,  
have a norm given by 
\beq
\langle \psi_{E_m} |  \psi_{E_m} \rangle =
 \int_{- \infty}^\infty \frac{d \om}{2 \pi}
 \; \frac{ |B(E_m) \ha(-\om) - A(E_m)  \hb( - \om)|^2}{( \om - E_m)^2}
\label{A9}
\eeq
In the examples discussed throughout   
the paper the functions $\ha(t), \hb(t)$
are phase factors, up to overall constants. Moreover, if the
function  $\ha(t) \hb(-t)$ is analytic in the upper half-plane
and vanishes when $|t| \rightarrow \infty, \	Re \;  t > 0$, then 
the $S$-functions and the associated Jost function take
a particular simple form if we allow for 
the existence of bound states, 
\beq
S_{a,a} = S_{b,b} = 1, \,  S_{a,b} = \ha(t) \; \hb(-t), 
\,  S_{b,a} = 0 \Longrightarrow
\F(t) = 2  + \ha(t) \; \hb(-t)
\label{A10}
\eeq
The  integration
constants $A,B$,  corresponding to a bound state,  can be choosen as 
\beq
A(E_m) = - B(E_m) = -1 
\label{A11}
\eeq
which differ with respect to (\ref{u29}) in an unimportant overall  sign.  
The wave function (\ref{A5}) also simplifies 
\beq
\psi_{E_m} (x)   = 
 \int_{-\infty}^{\infty} \frac{d \om}{2 \pi i} \; x^{-1/2 + i \om} \; 
\frac{ \ha(-\om) + \hb( - \om)}{ \om - E_m}
\label{A12}
\eeq
and  scalar product of two bound state wave functions becomes
\beq
\langle \psi_{E_{m_1} }|  \psi_{E_{m_2}} \rangle =
 \int_{- \infty}^\infty \frac{d \om}{2 \pi}
 \; \frac{ \F(\om) + \F(-\om)}{( \om - E_{m_1}) ( \om - E_{m_2}) }
\label{A13}
\eeq
The  analiticity of the Jost function $\F(E)$ in the upper-half plane 
implies the dispersion relation
\beq
\F(E) = \F_\infty + \int_{-\infty}^{\infty} \frac{d \om}{\pi i} \; 
\frac{ \F(\om)}{ \om - E_m}
\label{A14}
\eeq
where $\F_\infty$ is the value of $\F(E)$ at $E = + i \infty$. 
From this equation, and the fact that $\F(E_{m_1} ) = \F(E_{m_2} )=0$, 
one can show that $\psi_{E_{m_1} }$
and  $\psi_{E_{m_1} }$ are orthogonal. 
Furthermore,   eq.(\ref{A14}) yields also a simple expression for the
norm of  $\psi_{E_{m} }$
\beq
\langle \psi_{E_{m} }|  \psi_{E_{m}} \rangle =
 \int_{- \infty}^\infty \frac{d \om}{\pi}
 \; \frac{ \Re \F(\om) }{( \om - E_{m})^2 } = - \Im \; \F'(E_m) 
\label{A15}
\eeq
Finally, writing $\F(E)$ as in eq.(\ref{q38}), i.e.
\beq
\F(E) = 2 ( 1 + \ep \;  e^{2 \pi i n(E)} )
\label{A16}
\eeq
where $n(E)$ is the number of states, up to a constant, one
derives that the norm of $\psi_{E_{m} }$ is proportional
to the density of states at $E_m$, 
\beq
\langle \psi_{E_{m} }|  \psi_{E_{m}} \rangle =
4 \pi n'(E_m) 
\label{A17}
\eeq
\subsection{Wave functions associated to the smooth and exact 
Riemann zeros}

\begin{figure}[t!]
\begin{center}
\hbox{\includegraphics[height= 3.5 cm]{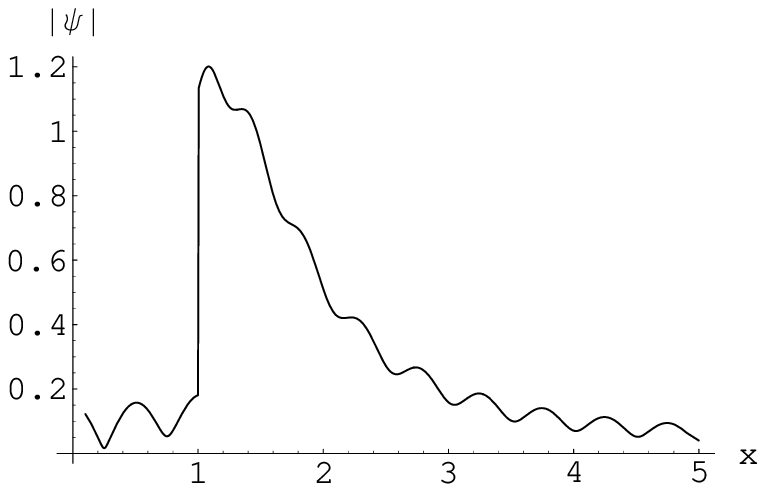}
\includegraphics[height = 3.5 cm]{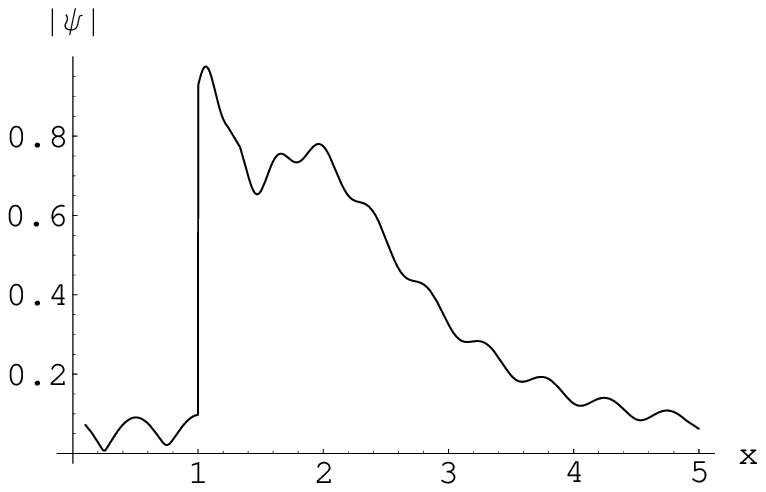}
\includegraphics[height = 3.5 cm]{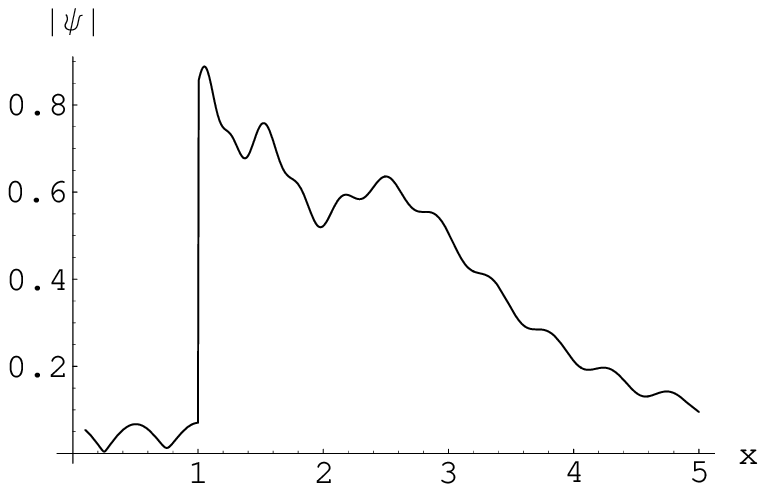}
}
\end{center}
\caption{Plot of $|\psi_{E_m}|$ for the energies $E_m= 14.5179,   20.654, 
25.4915$, 
corresponding to the lowest smooth Riemann zeros
(see eq.(\ref{A20})). The wave function are normalized
using eq.(\ref{A17}). 
}
\label{psi_bk}
\end{figure}

The Mellin transforms of the boundary wave functions
associated to the smooth Riemann zeros 
were given in eq.(\ref{r10}). Choosing $l_x = 1, l_p = 
2 \pi, a_0=b_0= \sqrt{2}$
we have 
\beq
\ha(t) = \sqrt{2} e^{2 i \theta(t)}, \qquad \hb(t) = 1
\label{A18}
\eeq
The wave functions (\ref{A12}) become in this case,
\beq
\psi_{E_m} (x)   = 
 \int_{-\infty}^{\infty} \frac{d \om}{\sqrt{2}  \pi i} \; x^{-1/2 + i \om} \; 
\frac{  e^{- 2 i \theta(\om)}+ 1 }{ \om - E_m}
\label{A19}
\eeq
 The integrals can be performed using the residue theorem
 obtaining 
 \beq
\frac{1}{\sqrt{2}} \psi_{E_m} (x)   = 
 \frac{H(x-1)}{x^{1/2 - i E_m}} + \frac{1}{\frac{1}{4} - 
\frac{i E_m}{2} } 
 \; _1F_2( \frac{1}{4} -  \frac{i E_m}{2} ;  \frac{1}{2}, 
\frac{5}{4} -  \frac{i E_m}{2},
 - \pi^2 x^2)
 \label{A20}
\eeq
 where $H(x-1) = 1$ if $x>1$ and 0 if $0<x<1$. One can show that 
 $\sqrt{x} \psi_{E_m} \rightarrow 0$ as $x \rightarrow \infty$, if 
 $1 + e^{2 i \theta(E_m)} =0$. 
 In fig.\ref{psi_bk} we plot the absolute values of (\ref{A20}) 
 for those energies that correspond to the three lowest Riemann zeros. 
 Notice that the functions are very small in the classical
forbidden region $0 < x < 1$. The amplitude has 
a high frequency component common to the three waves 
plus a low frequency one that depends on the level.

\begin{figure}[t!]
\begin{center}
\hbox{\includegraphics[height= 3.5 cm]{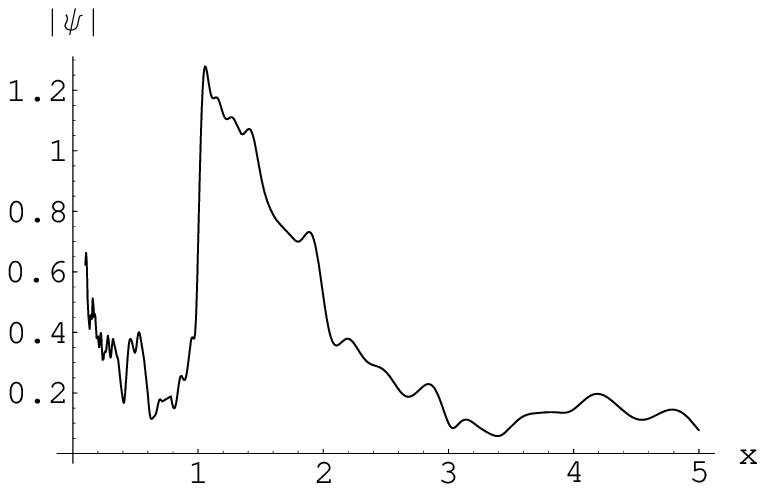}
\includegraphics[height = 3.5 cm]{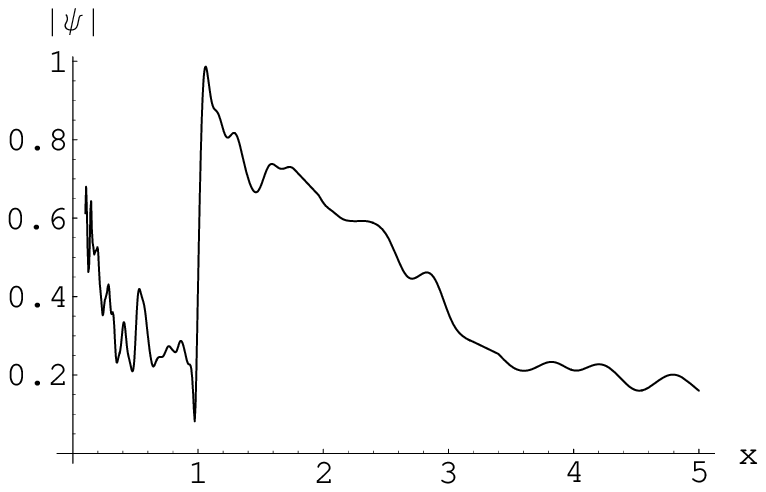}
\includegraphics[height = 3.5 cm]{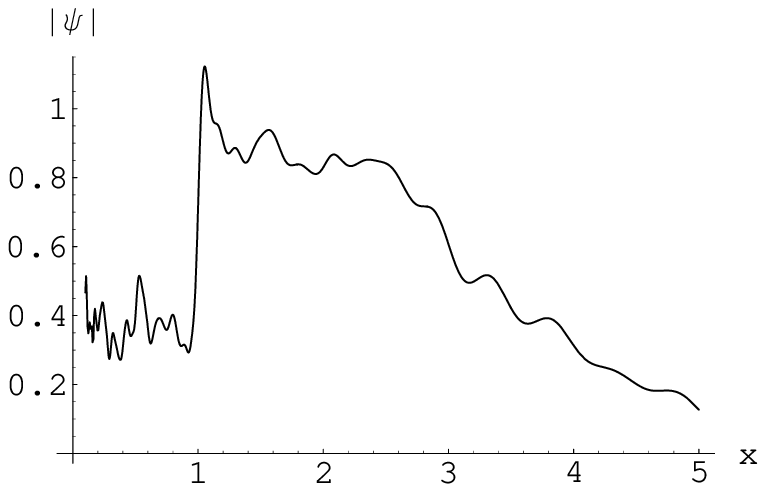}
}
\end{center}
\caption{Plot of $|\psi_{E_m}|$ for the energies 
Riemann zeros: 14.1347, 
21.022, 25.0109 evaluated with eq.(\ref{A21})
for $\Lambda = 60$.  The wave function are normalized
using eq.(\ref{A17}). 
}
\label{psi_r}
\end{figure}

\no
The wave functions associated to the exact Riemann 
zeros can be computed from eq.(\ref{A12}) 
with $\ha(t)$ and $\hb(t)$ given by eq. (\ref{q41}). 
We do not have an analytic expression for this
integral, however a numerical estimate can be obtained
truncating (\ref{A12}) as 
\beq
\psi_{E_m} (x)   \sim 
 \int_{E_m - \Lambda}^{E_m + \Lambda} 
\frac{d \om}{2 \pi i} \; x^{-1/2 + i \om} \; 
\frac{ \ha(-\om) + \hb( - \om)}{ \om - E_m}
\label{A21}
\eeq
\no
In fig.\ref{psi_r}  we plot the result for the lowests Riemann zeros. 
The wave functions have some common features with those
of fig. \ref{psi_bk}, but they also exhibit a random
behaviour.

\end{document}